\documentclass[journal=jpcafh,manuscript=article,longbibliography]{achemso}
\SectionNumbersOn
\usepackage[T1]{fontenc}

\author{Lucien Dupuy}
\affiliation[LCQS]{Laboratoire de Chimie Quantique,
Institut de Chimie, CNRS / Universit\'{e} de Strasbourg,
4 rue Blaise Pascal, 67000 Strasbourg, France}
\alsoaffiliation[usias]
{University of Strasbourg Institute for Advanced Study,
5, all\'{e}e du G\'{e}n\'{e}ral Rouvillois, F-67083 Strasbourg, France}
\email{lucien.dupuy@unistra.fr}
\author{Emmanuel Fromager}
\affiliation[LCQS]{Laboratoire de Chimie Quantique,
Institut de Chimie, CNRS / Universit\'{e} de Strasbourg,
4 rue Blaise Pascal, 67000 Strasbourg, France}
\alsoaffiliation[usias]
{University of Strasbourg Institute for Advanced Study,
5, all\'{e}e du G\'{e}n\'{e}ral Rouvillois, F-67083 Strasbourg, France}

\title
  {Exact static linear response of excited states from ensemble density functional theory
   }

\usepackage[colorlinks=true,urlcolor=blue,citecolor=blue,linkcolor=blue]{hyperref}

\usepackage{amsmath,bm}
\usepackage{amstext}
\usepackage{epsfig}
\usepackage{xcolor}
\usepackage{subfig}
\usepackage{graphicx,amsmath,amssymb,tabularx}
\usepackage{multirow}
\usepackage{array}
\usepackage{dsfont}
\usepackage{caption}
\usepackage{cleveref}
\usepackage{ulem}
\usepackage[toc]{appendix}
\usepackage{physics}
\usepackage{mathrsfs}  
\usepackage{amsbsy}

\definecolor{dgreen}{rgb}{0,.5,0}
\definecolor{dred}{rgb}{.7,.0,.0}
\definecolor{dblue}{RGB}{11, 139, 230}

\DeclareMathAlphabet\mathbfcal{OMS}{cmsy}{b}{n}

\newcommand{\bxi}{\bm{\xi}}

\newcommand{\br}{\mathbf{r}}

\newcommand{\ie}{{\it i.e.}}
\newcommand{\myket}[1]{\left\vert #1\right\rangle}
\newcommand{\mybra}[1]{\left\langle #1\right\vert}

\newcommand{\innerop}[3]{\left\langle #1 \middle\vert #2 \middle\vert #3 \right\rangle}
\newcommand{\contract}[2]{\left( #1 \middle\vert #2 \right)}

\newcommand{\be}{\begin{eqnarray}}
\newcommand{\ee}{\end{eqnarray}}
\newcommand{\bse}{\begin{subequations}}
\newcommand{\ese}{\end{subequations}}
\DeclareMathAlphabet\mathbfcal{OMS}{cmsy}{b}{n}

\newcommand{\rev}[1]{{{#1}}}

\begin{document}

\begin{abstract}
Following a recent work [E. Fromager, \textit{J. Phys. Chem. A} \textbf{2025}, \textit{129, 4}, 1143-1155] on the ensemble density functional theory (DFT) of excited electronic energy levels, we derive in this paper the ensuing static linear response theory, thus allowing for an in-principle exact evaluation of excited-state density-density linear response functions in a completely frequency-independent setting. Once individual-state components of the inverse ensemble linear response function have been introduced, a working Dyson-type equation naturally emerges for each state, individually. By considering the zero-weight limit of the theory, which infinitesimally deviates from standard Kohn--Sham DFT, exact excited-state corrections to ground-state linear response DFT can be identified. They involve the first-order weight derivatives of the ensemble Hartree-exchange-correlation (Hxc) potential and kernel, thus confirming the importance in ensemble DFT of both weight and density-functional derivatives of the ensemble Hxc energy functional.
\end{abstract}

\section{Introduction}\label{sec:introduction}

The advent of Density Functional Theory (DFT) was paramount in shaping the fields of quantum chemistry and material science into their present state~\cite{Teale2022_DFT_exchange}. 
Deducing all ground-state properties of electrons exactly from a universal functional of their density (whose explicit expression is unknown) brings a substantial reduction in complexity over wavefunction-based formalisms~\cite{hktheo}. 
The introduction of a fictitious, non-interacting (so-called Kohn-Sham (KS)) electronic system matching the ground-state density of the interacting one turned it into a practical tool~\cite{KS}: One can recover properties of a many-body system from greatly simplified calculations on its KS analog, with corrections derived from the Hartree-exchange-correlation (Hxc) density functional.
The latter needs to be approximated, and very successful approximate functionals have been introduced and refined over decades\cite{perdew1996generalized,Burke2012perspectiveDFA,Goerigk2017dftzoo}. 
Thanks to KS-DFT's unprecedented balance between accuracy and computational cost, ground-state electronic structure calculations have become feasible even for systems with thousands of atoms\cite{Tanaka2024largescaleDFT}.
The situation is however very different when one has to consider excited states' properties, as when studying photochemistry, charge transport etc. Indeed, there is no trivial extension to the excited states of the Hohenberg--Kohn (HK) theorem\cite{hktheo,Gorling99_Density,yang2024foundationdeltascfapproachdensity}, on which regular DFT relies, simply because excited-state energies are not (unconstrained) minima of the energy. For neutral excitation energies the most widely-used alternative is the time-dependent extension of DFT in the linear response regime~\cite{runge1984density,casida1995timedependent,Casida_tddft_review_2012,Lacombe2023_Non-adiabatic} (LR-TDDFT), which suffers from several limitations: The computational cost of solving the resulting Casida equation is substantially higher than that of a KS-DFT calculation. Moreover, when paired with the adiabatic approximation as most often done in practice, states with a double excitation character are completely missed from the KS spectrum which results in unphysical predictions~\cite{maitra2004double,cave2004dressed,Huix-Rotllant2011_Assessment,elliott2011perspectives,maitra2022double,Casida_tddft_review_2012,Lacombe2023_Non-adiabatic}. Another issue arises when one approaches the vicinity of a conical intersection between ground and first excited state\cite{Matsika2021elstructCI} as the single-reference character of the approach breaks down in presence of (quasi-)degeneracy of the former~\cite{Casida_tddft_review_2012}. This makes the development of cheap yet reliable methods for electronic structure of excited states an outstanding problem. Efforts are undertaken within the time-dependent framework\cite{Dar2023_TDDFT_DE,Dar2025_DTDDFT_CI,baranova2025excitedstatedensitiestimedependent} while alternative
time-independent DFTs of excited states have been developed
over the years at
both formal and practical
levels, either from an ensemble
perspective~\cite{Fan1949_On,JPC79_Theophilou_equi-ensembles,Hendekovic1982_equi-ensembles,gross1988rayleigh, gross1988density,oliveira1988density,deur2017exact,yang2017direct,gould2017hartree,gould2018charge,deur2018exploring,
PRL19_Gould_DD_correlation,Fromager_2020,PRL20_Gould_Hartree_def_from_ACDF_th,
Gould2020_Approximately,
loos2020weightdependent,Yang2021_Second,Gould2021_Ensemble_ugly,gould2024local,Gould2023_Electronic,gould2022single,
Gould2021_Double,
Cernatic2022,Schilling2021_Ensemble,Liebert2022_Foundation,Benavides-Riveros2022_Excitations,
Liebert_2023_An_exact_bosons,Liebert2023_Deriving,ding2024ground,Scott2024_Exact,Cernatic2024_Neutral,cernatic2024extended_doubles,Gould2025_PRL_tate-Specific} or a state-specific
one~\cite{Gorling99_Density,ayers2009pra,Gilbert08_Self-Consistent_MOM,Besley09_Self-consistent_MUM,Barca18_Simple_MOM,JCP09_Ziegler_relation_TD-DFT_VDFT,JCTC13_Ziegler_SCF-CV-DFT,levy2016computation,PRA12_Nagy_TinD-DFT_ES,JCP15_Ayers_KS-DFT_excit-states_Coulomb,Ayers2018_Article_Time-independentDensityFunctio,Garrigue2022,Giarrusso2023_Exact,Loos2025excistspHD,Loos2025excstUEG}. 
Alternate computational strategies like orbital-optimized DFT of excited states~\cite{Levi20_Variational,Ivanov21_Method,Hait21_Orbital,Schmerwitz22_Variational,OODFTRydberg_Levi2023,LeviOODFT_diamond_2023,Levi_OODFTvsTDDFT2024} have also shown very promising results. At this point we should stress (because it will be central in the rest of the paper) that, even though the Raleigh--Ritz variational principle does not apply to excited states, individually, the latter are stationary points of the energy, a property that can be exploited when rationalizing the aforementioned approaches from an ensemble perspective~\cite{gould2025stationaryconditionsexcitedstates,Fromager2025indvElevel}, for example. We adopt the latter perspective in the present work, thus allowing for an-principle exact density-functional description of excited-state properties.

Ensemble KS-DFT (EDFT)~\cite{Cernatic2022,gould2025ensemblizationdensityfunctionaltheory} constitutes a promising way of addressing the excited states’ electronic structure problem both from a formal and a practical standpoint. As an exact generalization of the KS-DFT setting to an ensemble of ground and excited states, it clarifies where the usual density mapping breaks down~\cite{Fromager_2020} and give explicit expressions of exact correction terms to the KS energy levels~\cite{deur2019ground}. This also applies to the analysis of {\it ad-hoc} schemes when they can be related to taking clearly defined approximations of the EDFT formalism\cite{gould2025stationaryconditionsexcitedstates,Fromager2025indvElevel}. As a computational tool, EDFT recovers the cheap framework of KS-DFT which is more affordable than LR-TDDFT and additionally does not suffer from the limitations of the latter regarding degeneracies\cite{Ullrich2001EDFTdegen}, double excitations\cite{marut2020weight,Gould2021_Double,sagredo2018can,cernatic2024extended_doubles} etc. As the core principle of EDFT only ensures the Raleigh--Ritz variational principle holds for an ensemble of states~\cite{gross1988rayleigh}, extraction of individual state properties is however a post-KS-DFT treatment. Several procedures have been proposed, among them are combining EDFT calculations for different ensembles\cite{gross1988density}, considering the ground-state limit of the theory\cite{yang2017direct,sagredo2018can,gould2022single,Amoyal_2025_pEDFT} or exploiting linearity in the ensemble weights from a single calculation~\cite{deur2019ground}. Strikingly, even recycling regular (weight-independent) density functional approximations (DFA) gave promising results, the correct weight-dependence of ensemble energy being for example recovered by linear interpolation between ground-state and equiensemble predictions\cite{senjean2015linear}. In the end, none of these extraction procedures are variational\cite{Fromager_2020}, in the sense of the Rayleigh--Ritz principle. More knowledge of the individual state properties regarding stationarity under certain conditions could prove fundamental in design of new approaches as it could allow to devise self-consistent iterative methods which were so successful in the original KS setting~\cite{Woods_2019scfdft} and extended in orbital-optimized DFT for excited states\cite{Levi20_Variational,Levi_DOMOM_2020,Ivanov21_Method,OODFT_GM_2023,schmerwitz2025freezeandreleasedirectoptimizationmethod}. Efforts in that direction have been undertaken by Gould~\cite{gould2025stationaryconditionsexcitedstates} through ensemble potential functional theory (EPFT) and by one of the authors in Ref.~\citenum{Fromager2025indvElevel} (that is referred to as Paper I from now on) by means of EDFT, showing that individual state energies can indeed be identified as ensemble density-functional stationary points. The individual-state stationarity condition within the ensemble led to a novel and still exact working equation to calculate excited-state densities from the KS ones through static ensemble density-functional linear response theory~\cite{Fromager2025indvElevel}. Moreover, it shed a new light on orbital-optimized DFT of excited states as an ensemble density functional approximation. In the present paper we extend the theory of Paper I to the evaluation of excited-state linear response static properties such as the electric polarizability, for example. An appealing feature of such a program is the fact that, unlike in the standard DFT+LR-TDDFT procedure, from which excited-state energies are computed and then differentiated with respect to the perturbation strength, the proposed strategy relies on a single and general frequency-independent formalism. \rev{In a practical mindset, excited state response functions allow to study change of polarizability\cite{LRMCSCF1985} of molecules after excitation, which is responsible for different chromophore behaviors depending on the solvent surrounding them\cite{C7CP08549D}.} To the best of our knowledge, the exact density-functional calculation of individual (ground or excited) state static responses within an ensemble has not been explored so far. As shown in the following, even if feasible in principle, this task is not straightforward and it raises various fundamental and practical questions that will be addressed in this work.

The Paper Is organized as follows: A brief summary of EDFT together with relevant findings of Paper I~\cite{Fromager2025indvElevel} is laid out in Sec.~\ref{sec:review_eDFT}. Then, we derive in Sec.~\ref{sec:results} an exact expression for the static linear response function of any individual (ground or excited) state within a given ensemble. The identification of a Dyson-type equation in this context is the key formal result of this section. Its various fundamental and practical implications are discussed in Sec.~\ref{sec:discussion}, with a particular focus on the definition of Hxc kernels for excited states (Sec.~\ref{sec:ind_pot_kernels}), the comparison in Sec.~\ref{sec:zero_weight_limit} of pure ground-state linear response DFT with the zero-weight limit of the theory (that deviates infinitesimally from the latter), and, finally, the perspective of constructing DFAs for ensembles by learning from the ground-state response within the ensemble (Sec.~\ref{sec:learning_from_GS_within_ensemble}). Our conclusions are drawn in Sec.~\ref{sec:conclusions}.

\section{Method}\label{sec:review_eDFT}

\subsection{TGOK ensemble DFT}

Theophilou--Gross--Oliveira--Kohn (TGOK) ensemble
DFT~\cite{JPC79_Theophilou_equi-ensembles,gross1988rayleigh, gross1988density,oliveira1988density,Cernatic2022} is an in-principle exact state-averaged flavor of DFT. Its basic variable is the state-averaged (so-called ensemble) density,
\be\label{eq:ens_dens_intro}
n^{\bxi}(\br):= \sum_{\nu\geq 0}\xi_\nu\,n_{\Psi_\nu}(\br)= n_{\Psi_0}(\br)+\sum_{\lambda>0}\xi_{\lambda}\left(n_{\Psi_\lambda}(\br)-n_{\Psi_0}(\br)\right),
\ee
where $n_\Psi(\br)\equiv \innerop{\Psi}{\hat{n}(\br)}{\Psi}$ denotes the electron density at position $\br$ of the many-electron wavefunction $\Psi$ and $\Psi_\nu$ is the $\nu$th eigenfunction of the electronic Hamiltonian operator,
\be\label{eq:phys_Hamil}
\hat{H}=\hat{T}+\hat{W}_{\rm ee}+\hat{V}_{\rm ext},
\ee
$\hat{T}$ being the kinetic energy operator, $\hat{W}_{\rm ee}$ the electron-electron
repulsion energy operator, and $\hat{V}_{\rm ext}=\int d\br\, v_{\rm
ext}(\br)\,\hat{n}(\br)$ the external local potential operator. In Eq.~(\ref{eq:ens_dens_intro}), $\bxi\equiv \left\{\xi_\lambda\right\}_{\lambda>0}$ denotes the collection of {\it independent} ensemble weights that are assigned to the (neutral in the present context) excited states ($\nu=\lambda>0$). The weight assigned to the ground state $(\nu=0)$,    
\be\label{eq:ens_normalization}
\xi_0=1-\sum_{\lambda>0}\xi_\lambda,
\ee
ensures that the ensemble density integrates to the correct and fixed (integer) number of electrons. Under the assumption that the weights are ordered as follows,
\be\label{eq:GOK_weights_ordering}
0\leq \xi_{\nu+1}\leq \xi_\nu,\;\;\nu\geq 0,
\ee
the exact ensemble energy
\be\label{eq:def_true_ens_ener}
E^{\bxi}:=\sum_{\nu\geq 0}\xi_\nu\,E_\nu,
\ee
$E_\nu$ being the $\nu$th eigenvalue of $\hat{H}$, can be determined {\`a} la Kohn--Sham as follows,
\begin{subequations}\label{eq:ens_dens_VP}
\begin{align}
E^{\bxi} &= \min_n \left\{ F^{\bxi}[n]+\left(v_{\rm ext}\vert n\right) \right\} 
\\
\label{eq:min_dens_ens_KS-DFT}
&=\min_n\left\{T_{\rm s}^{\bxi}[n]+E^{\bxi}_{\rm
Hxc}[n]+\contract{v_{\rm ext}}{n}\right\}, 
\end{align}
\end{subequations}
where $\left(v_{\rm ext}\vert n\right):=\int
d\br\,v_{\rm ext}(\br)\,n(\br)$ and the minimizing density is the exact ensemble one $n^{\bxi}$. The analog for ensembles of the HK~\cite{hktheo,gross1988density} $F^{\bxi}[n]$ and non-interacting kinetic energy $T_{\rm s}^{\bxi}[n]$ functionals read 
\be\label{eq:ens_HK_func_def}
F^{\bxi}[n]= \sum_{\nu\geq 0}\xi_\nu\,\mybra{\Psi^{{\bxi}}_\nu[n]}\hat{T}+\hat{W}_{\rm
ee}\myket{\Psi^{{\bxi}}_\nu[n]}
\ee
and
\be\label{eq:ens_Ts_func_def}
T_{\rm s}^{\bxi}[n]=\sum_{\nu\geq 0}\xi_\nu\,T^{{\bxi}}_{{\rm
s},\nu}[n]=\sum_{\nu\geq 0}\xi_\nu\,\mybra{\Phi^{{\bxi}}_\nu[n]}\hat{T}\myket{\Phi^{{\bxi}}_\nu[n]},
\ee
respectively. Note that the Hxc energy $E^{\bxi}_{\rm
Hxc}[n]=F^{\bxi}[n]-T_{\rm s}^{\bxi}[n]$ of the many-body ensemble is a functional of the density {\it and} a function of the ensemble weights, both being {\it independent} variables~\cite{Cernatic2022,Fromager2025indvElevel}. Developing weight-dependent density-functional approximations for ensembles still remains a challenging task, even though substantial progress has been made recently~\cite{Gould2025_PRL_tate-Specific}. In this work we mainly focus on the exact theory and the decomposition of $E^{\bxi}_{\rm
Hxc}[n]$ into separate Hartree, exchange, and correlation contributions will not be considered. We refer the reader to Refs.~\citenum{Cernatic2022} and ~\citenum{gould2025ensemblizationdensityfunctionaltheory}, and the references therein, for further details on this (non-trivial) point in the context of ensemble DFT. Returning to the density functionals of Eqs.~(\ref{eq:ens_HK_func_def}) and (\ref{eq:ens_Ts_func_def}), the interacting $\left\{\Psi^{{\bxi}}_\nu[n]\right\}$ and non-interacting $\left\{\Phi^{{\bxi}}_\nu[n]\right\}$ ensemble density-functional wavefunctions are eigenfunctions of $\hat{T}+\hat{W}_{\rm
ee}+\int d\br\;v^{\bxi}[n](\br)\hat{n}(\br)$ and $\hat{T}+\int d\br\;v_{\rm s}^{\bxi}[n](\br)\hat{n}(\br)$, respectively, where the interacting $v^{\bxi}[n]$ and non-interacting KS $v_{\rm s}^{\bxi}[n]$ ensemble density-functional potentials are such that the following ensemble density constraint is fulfilled: \be\label{eq:int_ens_dens_constraint}
\sum_{\nu\geq 0}\xi_\nu\,n_{\Psi^{{\bxi}}_\nu[n]}=n=\sum_{\nu\geq 0}\xi_\nu\,n_{\Phi^{{\bxi}}_\nu[n]}.
\ee
Like in regular ground-state DFT, these two potentials can be expressed as density-functional derivatives~\cite{Fromager2025indvElevel}, \ie,
\be\label{eq:int_ens_dens_func_pot_as_derivative}
v^{\bxi}[n]=-\dfrac{\delta F^{\bxi}[n]}{\delta n} 
\ee
and
\be\label{eq:KS_ens_dens_pot_exp_part1}
v_{\rm s}^{\bxi}[n]=-\dfrac{\delta T_{\rm s}^{\bxi}[n]}{\delta
n},
\ee
thus leading to the key relation between the interacting and non-interacting KS ensembles, 
\be\label{eq:KS_ens_dens_pot_exp}
v_{\rm s}^{\bxi}[n]=v^{\bxi}[n]+v^{\bxi}_{\rm Hxc}[n],
\ee
where $v^{\bxi}_{\rm Hxc}[n]={\delta E^{\bxi}_{\rm Hxc}[n]}/{\delta n}$ is the Hxc ensemble density-functional potential. When written for the exact ensemble density it gives 
\be\label{eq:true_ens_KS_pot}
v_{\rm s}^{{\bxi}}:=\left.v_{\rm
s}^{{\bxi}}[n]\right|_{n=n^{\bxi}}=v_{\rm ext}+v^{\bxi}_{\rm
Hxc}[n^{\bxi}],
\ee
so that the targeted KS wavefunctions $\left\{\Phi^{{\bxi}}_\nu:=\Phi^{{\bxi}}_\nu\left[n^{\bxi}\right]\right\}$, which are all constructed from the same set of orthonormal ensemble KS orbitals $\left\{\varphi_i^{\bxi}\right\}$, can be determined by solving the self-consistent ensemble KS equations,  
\be
\left(-\dfrac{\nabla^2_{\br}}{2}+v_{\rm ext}+v^{\bxi}_{\rm
Hxc}[n^{\bxi}]\right)\varphi_i^{\bxi}(\br)=\varepsilon_i^{\bxi}\varphi_i^{\bxi}(\br),
\ee
where 
\begin{subequations}
\begin{align}
n^{\bxi}(\br)=\sum_{\nu\geq 0}\xi_\nu\,n_{\Phi^{{\bxi}}_\nu}(\br)
&\equiv \sum_{\nu\geq 0}\xi_\nu\left(\sum_{i}\theta^{\nu}_i\left|\varphi_i^{\bxi}(\br)\right|^2\right)
\\
\label{eq:KS_ens_dens_frac_occ_orbs}
&= \sum_{i}\left(\sum_{\nu\geq 0}\xi_\nu\theta^{\nu}_i\right)\left|\varphi_i^{\bxi}(\br)\right|^2
,
\end{align}
\end{subequations} 
$\theta^{\nu}_i$ denoting the occupancy of the orbital $\varphi_i^{\bxi}$ in $\Phi^{{\bxi}}_\nu$. As readily seen from Eq.~(\ref{eq:KS_ens_dens_frac_occ_orbs}), unlike in regular KS-DFT, the occupation of the ensemble KS orbitals can be fractional and it is controlled by the ensemble weights. 

\subsection{Linear response ensemble DFT}

Our previous TGOK ensemble DFT summary is laid out assuming a given external potential $v_{\text{ext}}$ defining the electronic Hamiltonian $\hat{H}$ together with the number of electrons. We emphasize the theory is valid for any (one-body) local potential $v$. It is thus more general to view the latter as a {\it parameter} alongside the set of ensemble weights $\bxi$, so that, for example, the ensemble energy stationarity condition (that originates from the minimization in Eq.~(\ref{eq:min_dens_ens_KS-DFT})) takes the more explicit form:
\begin{equation} \label{eq:statio_ense_anyv}
     \frac{\delta E_v^{\bxi}[n]}{\delta n}\bigg|_{n=n_v^{\mathbf{\bxi}}} =  0,\;\;\forall v,    \end{equation}
where
\be \label{eq:ens_dens_VP_v}
E_v^{\bxi}[n]=T_{\rm s}^{\bxi}[n]+E^{\bxi}_{\rm
Hxc}[n]+\contract{v}{n}.
\ee
When $v=v_{\text{ext}}$, the true ensemble density $n_{v_{\text{ext}}}^{\bxi}=n^{\bxi}$ of the system under study is recovered. As the stationarity condition above enables to retrieve that density, in principle exactly, from a non-interacting KS ensemble~\cite{Fromager2025indvElevel}, considering an additional variation with respect to $v$ shall offer insight into the (static) linear response of the many-body ensemble:
\be
n_{v_{\rm ext}+\delta v}^{\bxi}(\br)=n^{\bxi}(\br)+\int d\br'\;\chi^{\bxi}(\br,\br')\delta v(\br'),
\ee
that we write in a more compact way as follows,
\be
n_{v_{\rm ext}+\delta v}^{\bxi}=n^{\bxi}+\chi^{\bxi}\star\delta v,
\ee
$\chi^{\bxi}$ being the true physical ensemble (density-density) linear response function, \ie,
\be
\chi^{\bxi}=\left.\dfrac{\delta n_v^{\bxi}}{\delta v}\right|_{v=v_{\text{ext}}}.
\ee
The purpose of the present work is to extract (from the response of the ensemble) the individual response of any (ground or excited) state. This will be achieved in Sec.~\ref{sec:results} by considering variations in $v$ of individual ensemble density-functional energy levels. As a warm-up exercise, we here consider the simpler case of ensemble linear response theory emerging from:
\begin{equation} 
    \left.\frac{\delta}{\delta v}\left[ \frac{\delta E_v^{\bxi}[n]}{\delta n}\Bigg|_{n=n_v^{\mathbf{\bxi}}}  \right] \right|_{v=v_{\text{ext}}} =  0,    
\end{equation}
which follows immediately from Eq.~\eqref{eq:statio_ense_anyv}.  Inserting Eq.~\eqref{eq:ens_dens_VP_v} and evaluating the inner derivative using Eqs.~(\ref{eq:int_ens_dens_func_pot_as_derivative},\ref{eq:KS_ens_dens_pot_exp_part1}) yields
\begin{equation}\label{eq:ens_LR_eq_deriv_v}
    \frac{\delta}{\delta v}\left[ v + v^{\bxi}_{\text{Hxc}}\big[n^{\bxi}_v\big] - v^{\bxi}_{\rm s}\big[n^{\bxi}_v\big] \right] \Big|_{v=v_{\text{ext}}} = 0.
\end{equation}
Before evaluating the remaining derivative, let us recall, for the sake of clarity, the principle of chain rule applied to functional differentiation:
\be
\frac{\delta n(\br)}{\delta n(\br')}=\delta\left(\br-\br'\right)=\int d\br''\dfrac{\delta n(\br)}{\delta v_{\rm s}^{\bxi}[n](\br'')}\dfrac{\delta v_{\rm s}^{\bxi}[n](\br'')}{\delta n(\br')},
\ee
which can be rewritten in a more compact form as follows,
\be
\hat{1}=\chi_{\rm s}^{\bxi}[n]\star\dfrac{\delta v_{\rm s}^{\bxi}[n]}{\delta n},
\ee
or, equivalently,
\be\label{eq:inv_rsp_func_ens_KS}
\dfrac{\delta v_{\rm s}^{\bxi}[n]}{\delta n}=\left(\chi_{\rm s}^{\bxi}[n]\right)^{-1},
\ee
where $\chi_{\rm s}^{\bxi}[n]$ is the density-density linear response function of the density-functional KS ensemble. The $\star$ notation introduced above will be used extensively to present more involved chain rules performed in the following. Returning to Eq.~(\ref{eq:ens_LR_eq_deriv_v}), use of Eq.~(\ref{eq:inv_rsp_func_ens_KS}) together with the definition of the ensemble density-functional Hxc kernel
\begin{equation}
    f_{\text{Hxc}}^\xi[n] = \frac{\delta v_{\text{Hxc}}^\xi[n]}{\delta n}
\end{equation}
gives the compact form
\begin{equation}
    \hat{1} + f^{\bxi}_{\text{Hxc}} \star \chi^{\bxi} - (\chi^{\bxi}_{\rm s})^{-1} \star \chi^{\bxi}=\hat{1}+\left(f^{\bxi}_{\text{Hxc}}-(\chi^{\bxi}_{\rm s})^{-1}\right)\star\chi^{\bxi}=0,
\end{equation}
where all quantities are evaluated for $v=v_{\text{ext}}$ and $n=n^{\bxi}$. Thus we recover the ensemble Dyson equation (that was originally derived in Ref.\citenum{Fromager2025indvElevel} for any ensemble density $n$):
\begin{equation} \label{eq:ensdyson}
    (\chi^{\bxi})^{-1} = (\chi^{\bxi}_{\rm s})^{-1} - f^{\bxi}_{\text{Hxc}}.
\end{equation}
As shown later in Sec.~\ref{sec:results}, this equation is central but it will not be sufficient to study the response of excited states individually. We will need to derive an additional Dyson equation for that purpose.

\subsection{Ensemble DFT of individual energy levels and densities}

We recall in this section the main results of Paper I that will be used in Sec.~\ref{sec:results} to obtain individual linear response properties of any state within the ensemble. The first one is the exact KS-DFT-like ensemble density-functional expression of the $\nu$th energy level (degeneracies have not been considered for simplicity but can be treated by means of multiplets~\cite{gross1988density}, if necessary) for a given local potential $v$ and a choice of ensemble weights $\bxi$ (see Eq.~(56) of the original paper~\cite{Fromager2025indvElevel}): 
\begin{subequations} \label{eq:indvElvl}
    \begin{align}    
        E^{\bxi}_{v,\nu}[n] = & T^{\bxi}_{{\rm s},\nu}[n] + E^{\bxi}_{\text{Hxc}}[n] + \sum_{\lambda>0}(\delta_{\lambda\nu}-\xi_{\lambda}) \frac{\partial E^{\bxi}_{\text{Hxc}}[n]}{\partial\xi_\lambda}
        \\
        & - \left( \frac{\delta E^{\bxi}_{\text{Hxc}}[n]}{\delta n} \Bigg| n \right) + \left( \frac{\delta F^{\bxi}[n]}{\delta n} + v\bigg| n_{\Psi^{\bxi}_{\nu}}[n] \right)
        \\
        & - \left( \frac{\delta T_{\rm s}^{\bxi}[n]}{\delta n} \Bigg| n_{\Phi^{\bxi}_{\nu}}[n] \right),
    \end{align}
\end{subequations}
that we write equivalently (see Eqs.~(\ref{eq:int_ens_dens_func_pot_as_derivative}), (\ref{eq:KS_ens_dens_pot_exp_part1}), and (\ref{eq:KS_ens_dens_pot_exp})), for the purpose of deriving linear response equations later on, as follows,
\begin{subequations} \label{eq:indvElvl_in_terms_of_pots}
    \begin{align}    
        E^{\bxi}_{v,\nu}[n] = & T^{\bxi}_{{\rm s},\nu}[n] + E^{\bxi}_{\text{Hxc}}[n] + \sum_{\lambda>0}(\delta_{\lambda\nu}-\xi_{\lambda}) \frac{\partial E^{\bxi}_{\text{Hxc}}[n]}{\partial\xi_\lambda}
        \\
        & - \contract{v^{\bxi}_{\rm{Hxc}}[n]}{n} + \contract{v-v_{\rm s}^{\bxi}[n]+v^{\bxi}_{\rm{Hxc}}[n]}{n_{\Psi^{\bxi}_{\nu}}[n]}
        \\
        & +\contract{v_{\rm s}^{\bxi}[n]}{n_{\Phi^{\bxi}_{\nu}}[n]}.
    \end{align}
\end{subequations}
The second key result of Paper I is the stationarity of each energy level within the ensemble, 
\begin{equation} \label{eq:stindvElvl}
    \frac{\delta E^{\bxi}_{v,\nu}[n]}{\delta n} \Big|_{n=n_v^{\bxi}} = 0,\;\;\forall v,
\end{equation}
from which an exact relation between KS and true individual densities (that do not necessarily match, unlike the ensemble ones~\cite{PRL19_Gould_DD_correlation,Fromager_2020}) can be recovered (see Eq.~(66) of Ref.~\citenum{Fromager2025indvElevel}):
\begin{equation}\label{eq:true_dens_from_stat_cond}
n_{\Psi_\nu} - n_{\Phi_\nu^{\bxi}}
= \chi_{\rm s}^{\bxi}\star \sum_{\lambda>0}\left(\delta_{\lambda\nu}
-\xi_\lambda\right)\dfrac{\partial
v^{\bxi}_{\rm Hxc}}{\partial\xi_\lambda}.
\end{equation}
As pointed out in Paper I, the above relation can also be deduced from the linear variation of the exact ensemble density with the ensemble weights $\bxi$ (see Eq.~(\ref{eq:ens_dens_intro})), which implies (see Eq.~(73d) of Ref.~\citenum{Fromager2025indvElevel}): 
\begin{equation} \label{eq:npsinxi}
      n_{\Psi_\nu}-n^{\bxi}= \sum_{\lambda>0} (\delta_{\lambda\nu}-\xi_\lambda)  \frac{\partial n^{\bxi}}{\partial\xi_\lambda}.  
\end{equation} 
Paper I also provided the following working equation (see Eqs.~(76,83) in Ref.~\citenum{Fromager2025indvElevel}),
\be \label{eq:dnxi}
\dfrac{\partial n^{\bxi}}{\partial
\xi_\lambda}=\left(\hat{1}+\chi^{\bxi}\star f^{\bxi}_{\rm Hxc}\right)\star 
\Delta
n^{\bxi}_{{\rm s},\lambda},
\ee
where
\be\label{eq:gathering_KS_terms_deriv_ens_dens_wrt_weight}
\Delta
n^{\bxi}_{{\rm s},\lambda}
=\left(n_{\Phi_\lambda^{\bxi}}-n_{\Phi_0^{\bxi}}\right)+\chi^{\bxi}_{\rm
s}\star \left.\dfrac{\partial
v^{\bxi}_{\rm Hxc}[n]}{\partial
\xi_\lambda}\right|_{n=n^{\bxi}},
\ee
thus showing that the physical densities can indeed be evaluated individually and exactly from the ensemble Hxc functional and its weight-derivatives.

\section{Results}\label{sec:results}

We will now show how the stationarity of individual energy levels (see Eqs.~(\ref{eq:indvElvl},\ref{eq:stindvElvl})) exploited in Paper I\cite{Fromager2025indvElevel} allows to derive an equation for individual (density-density) linear response functions relating the true many-body and KS ensembles. We do so by considering implications of an additional variation with respect to the external potential, which also gives zero as the stationarity condition holds for {\it any} potential $v$ (that will infinitesimally deviate from $v_{\text{ext}}$ in the present context):
\begin{equation} \label{eq:dblevarEnu}
    \frac{\delta}{\delta v}\left[ \frac{\delta E^{\bxi}_{v,\nu}[n]}{\delta n}\Bigg|_{n=n^{\mathbf{\bxi}}_v}  \right] \Bigg|_{v=v_{\text{ext}}} =  0.    
\end{equation}
Inserting the exact ensemble density-functional energy level expression of Eq.~(\ref{eq:indvElvl_in_terms_of_pots}) into Eq.~(\ref{eq:dblevarEnu}) yields
\begin{subequations}
    \begin{align}
        0 = & \frac{\delta}{\delta v}\Bigg[ \frac{\delta}{\delta n}\bigg[ T^{\bxi}_{{\rm s},\nu}[n] + E^{\bxi}_{\text{Hxc}}[n] + \sum_{\lambda>0}(\delta_{\lambda\nu}-\xi_{\lambda}) \frac{\partial E^{\bxi}_{\text{Hxc}}[n]}{\partial\xi_\lambda} -  \contract{v^{\bxi}_{\text{Hxc}}[n]}{n}
        \\
        &   + \Big( - \big(v^{\bxi}_{\rm s}[n]-v^{\bxi}_{\text{Hxc}}[n]\big) + v \Big| n_{\Psi^{\bxi}_{\nu}}[n] \Big)
        + \left( v^{\bxi}_{\rm s}[n] \Big| n_{\Phi^{\bxi}_{\nu}}[n] \right)  \bigg]\bigg|_{n=n^{\mathbf{\bxi}}_v} \Bigg]\Bigg|_{v=v_{\text{ext}}},
    \end{align}
\end{subequations}
or, more explicitly (see Eq.~(\ref{eq:inv_rsp_func_ens_KS})),
\begin{subequations} \label{eq:dblsta1}
    \begin{align}
    0 = & \frac{\delta}{\delta v}\Bigg[ 
    \frac{\delta}{\delta n}\bigg[ T^{\bxi}_{{\rm s},\nu}[n] +\contract{v_{\rm s}^{\bxi}\left[n^{\mathbf{\bxi}}_v\right]}{n_{\Phi^{\bxi}_\nu[n]}}\bigg]_{n=n^{\mathbf{\bxi}}_v}\Bigg]\Bigg|_{v=v_{\text{ext}}}
    \\ 
    & +\frac{\delta}{\delta v}\Bigg[\sum_{\lambda>0} (\delta_{\lambda\nu}-\xi_\lambda) \left.\frac{\partial v^{\bxi}_{\text{Hxc}}[n]}{\partial\xi_\lambda}\right|_{n=n^{\mathbf{\bxi}}_v} - f^{\bxi}_{\text{Hxc}}\big[n^{\bxi}_v] \star n^{\bxi}_v 
    \\
    & + \big(\chi^{\bxi}_{\rm s} \big[n^{\bxi}_v\big]\big)^{-1} \star n_{\Phi_\nu^{\bxi}\left[n^{\bxi}_v\right]} - \left(\big(\chi^{\bxi}_{\rm s} \big[n^{\bxi}_v\big]\big)^{-1} - f^{\bxi}_{\text{Hxc}}\big[n^{\bxi}_v\big]  \right) \star n_{\Psi_\nu^{\bxi}\left[n^{\bxi}_v\right]}
    \\
    & + \left( v - \left(v^{\bxi}_{\rm s}\Big[n^{\bxi}_v\Big]-v^{\bxi}_{\text{Hxc}}\Big[n^{\bxi}_v\Big]\right) \bigg| \frac{\delta n_{\Psi_\nu^{\bxi}[n]}}{\delta n}\Big|_{n=n^{\bxi}_v} \right) \Bigg] \Bigg|_{v=v_{\text{ext}}}
    .
    \end{align}
\end{subequations}
Before carrying on (in Appendix~\ref{secapp:deriv_stat_cond}) with the explicit differentiation with respect to $v$, we can actually proceed to several simplifications: The first line on the right-hand-side of Eq.~(\ref{eq:dblsta1}) is zero, as a consequence of the ensemble KS equations (for a given $v$, KS wavefunctions are stationary solutions to the ensemble density-functional KS Hamiltonian~\cite{Fromager2025indvElevel}). In addition, the last line also cancels owing to the expression of the exact ensemble KS potential (see Eq.~(\ref{eq:true_ens_KS_pot})): 
\be\label{eq:KS_pot_for_any_local_v}
v^{\bxi}_{\rm s}\big[n^{\bxi}_v\big]:=v+v^{\bxi}_{\text{Hxc}}\big[n^{\bxi}_v\big].
\ee
As shown in detail in Appendix~\ref{secapp:deriv_stat_cond}, deriving the remaining non-zero contributions enables to relate the linear response function $\chi^{\bxi}_{{\rm s},\nu}$ of the $\nu$th state within the KS ensemble to that of the physical $\nu$th interacting state $\chi_\nu$ as follows,
\begin{subequations} \label{eq:dblsta2}
    \begin{align}
    \big(\chi^{\bxi}\big)^{-1} \chi_\nu = & \bigg[\sum_{\lambda>0} (\delta_{\lambda\nu}-\xi_\lambda) \frac{\partial f^{\bxi}_{\text{Hxc}}[n]}{\partial\xi_\lambda}\Bigg|_{n=n^{\bxi}}   + \left(g^{\bxi}_{\text{Hxc}} \star (n_{\Psi_\nu}-n^{\bxi})\right)  - f^{\bxi}_{\text{Hxc}} \bigg] \star \chi^{\bxi}
    \\
    & + \big(\chi^{\bxi}_{\rm s}\big)^{-1} \star \bigg[ \chi^{\bxi}_{{\rm s},\nu}  +  \left(\chi^{[2]\bxi}_{\rm s} \star \big(\chi^{\bxi}_{\rm s} \big)^{-1} \star (n_{\Psi_\nu}-n_{\Phi_\nu^{\bxi}})\right)  \bigg] \star \big(\chi^{\bxi}_{\rm s}\big)^{-1}  \star \chi^{\bxi},
    \end{align}
\end{subequations}
where, as readily seen, knowing the (three-point) quadratic response function $\chi^{[2]\bxi}_{\rm s}: (\br,\br',\br'')\mapsto \delta \chi^{\bxi}_{\rm s}(\br,\br')/\delta v^{\bxi}_{\rm s}(\br'')$  of the KS ensemble and the quadratic response ensemble (static) Hxc kernel $g^{\bxi}_{\text{Hxc}}: (\br,\br',\br'')\mapsto \left.\delta f^{\bxi}_{\rm Hxc}[n](\br,\br')/\delta n(\br'')\right|_{n=n^{\bxi}}$ is necessary, in principle, in addition to the physical ensemble linear response function $\chi^{\bxi}$. An explicit sum-over-states expression of $\chi^{[2]\bxi}_{\rm s}$ is derived in Appendix~\ref{app:SOS_KS_quadratic_rsp_func}, for completeness.\\

While Eq.~(\ref{eq:dblsta2}) makes our attempt to further exploit the stationarity of ensemble density-functional energy levels successful, it also reveals (as we could have expected) that, at least at first sight, exact excited-state linear response functions cannot be straightforwardly retrieved from a Dyson-like equation,
\begin{equation}
    \chi_\nu \neq \chi^{\bxi}_{{\rm s},\nu}+ \chi^{\bxi}_{{\rm s},\nu}\star f^{\bxi}_{\text{Hxc},\nu}\star\chi_\nu.
\end{equation}
In order words, no Hxc kernel $f^{\bxi}_{\text{Hxc},\nu}$ for the $\nu$th state of interest emerges naturally from Eq.~(\ref{eq:dblsta2}):
\be
f^{\bxi}_{\text{Hxc},\nu}\neq \left(\chi^{\bxi}_{{\rm s},\nu}\right)^{-1}-\left(\chi_\nu\right)^{-1}.
\ee

In fact, Eq.~(\ref{eq:dblsta2}) can be rephrased as a Dyson equation but, for that purpose, auxiliary quantities (from which linear response functions can be equivalently evaluated, as shown later in Eq.~(\ref{eq:true_lrfunct_final_expression_from_ens})) must be introduced. More precisely, we consider the following $\nu$th component of the inverse ensemble linear response function,
\begin{equation} \label{eq:Rnu}
    (R^{\bxi}_\nu)^{-1} =  \big(\chi^{\bxi}\big)^{-1} \star \chi_\nu \star \big(\chi^{\bxi}\big)^{-1},
\end{equation}
where, by construction (see Eq.~(\ref{eqapp:true_ens_rsp_func_exp})), 
\be\label{eq:weighted_sum_Rnu}
\sum_{\nu\geq0} \xi_\nu (R^{\bxi}_{\nu})^{-1} = \big(\chi^{\bxi}\big)^{-1} \star \chi^{\bxi} \star (\chi^{\bxi})^{-1}=\big(\chi^{\bxi}\big)^{-1},
\ee
hence the name given to $(R^{\bxi}_{\nu})^{-1}$. Its KS analog reads
\begin{equation}\label{eq:Rnu_KS}
    (R^{\bxi}_{{\rm s},\nu})^{-1} =  \big(\chi^{\bxi}_{\rm s}\big)^{-1} \star \chi^{\bxi}_{{\rm s},\nu} \star \big(\chi^{\bxi}_{\rm s}\big)^{-1},
\end{equation}
and, similarly,
\begin{equation}\label{eq:weighted_sum_Rnu_KS}
        \sum_{\nu\geq0} \xi_\nu (R^{\bxi}_{{\rm s},\nu})^{-1} =  (\chi^{\bxi}_{\rm s})^{-1}.
\end{equation}

From these definitions, it now becomes clear that Eq.~(\ref{eq:dblsta2}) can be rewritten as an {\it effective} Dyson equation,
\begin{equation} \label{eq:dysonlike}
    (R^{\bxi}_{\nu})^{-1} = (R^{\bxi}_{{\rm s},\nu})^{-1} -\Xi^{\bxi}_{{\rm Hxc},\nu},
\end{equation}
with the individual-state and weight-dependent kernel $\Xi^{\bxi}_{{\rm Hxc},\nu}$ being identified as
\begin{subequations} \label{eq:Xinu}
    \begin{align}
        \Xi^{\bxi}_{{\rm Hxc},\nu} = & -\sum_{\lambda>0} (\delta_{\lambda\nu}-\xi_\lambda) \frac{\partial f^{\bxi}_{\text{Hxc}}[n]}{\partial\xi_\lambda}\Bigg|_{n=n^{\bxi}} - \left(g^{\bxi}_{\text{Hxc}} \star (n_{\Psi_\nu}-n^{\bxi})\right)  + f^{\bxi}_{\text{Hxc}}
        \\
        & -  \big(\chi^{\bxi}_{\rm s}\big)^{-1} \star \bigg(  \chi^{[2]\bxi}_{\rm s} \star \big(\chi^{\bxi}_{\rm s} \big)^{-1} \star (n_{\Psi_\nu}-n_{\Phi_\nu^{\bxi}})  \bigg) \star \big(\chi^{\bxi}_{\rm s}\big)^{-1}. 
    \end{align}
\end{subequations}
Note that, by construction (see Eqs.~(\ref{eq:weighted_sum_Rnu}) and (\ref{eq:weighted_sum_Rnu_KS})), the above Hxc kernel is the $\nu$th component of the total ensemble Hxc kernel:
\be\label{eq:average_ind_kernels_Xi_def}
\sum_{\nu\geq0} \xi_\nu \Xi^{\bxi}_{{\rm Hxc},\nu}= (\chi^{\bxi}_{\rm s})^{-1}-\big(\chi^{\bxi}\big)^{-1}=f^{\bxi}_{\text{Hxc}}.
\ee
Eq.~(\ref{eq:Xinu}) (see also the fully expanded expression of Eq.~(\ref{eq:weXi})), combined with Eqs.~(\ref{eq:Rnu}), (\ref{eq:Rnu_KS}), (\ref{eq:dysonlike}), and the ensemble Dyson Eq.~(\ref{eq:ensdyson}), is the first key result of the present work. It is conceptually significant as the kernel $\Xi^{\bxi}_{{\rm Hxc},\nu}$ does not depend on $\chi^{\bxi}_{{\rm s},\nu}$ (nor $R^{\bxi}_{{\rm s},\nu}$). For completeness, note that, once the ensemble Eq.~(\ref{eq:ensdyson}) has been solved and $\Xi^{\bxi}_{{\rm Hxc},\nu}$ has been evaluated (see Eq.~(\ref{eq:Xinu})), the targeted excited-state linear response function can be calculated as follows,
\be
\chi_\nu=\chi^{\bxi}\star\left(\big(\chi^{\bxi}_{\rm s}\big)^{-1} \star \chi^{\bxi}_{{\rm s},\nu} \star \big(\chi^{\bxi}_{\rm s}\big)^{-1}-\Xi^{\bxi}_{{\rm Hxc},\nu}\right)\star\chi^{\bxi},
\ee
or, equivalently,
\be\label{eq:true_lrfunct_final_expression_from_ens}
\chi_\nu=\left(\hat{1}+\chi^{\bxi}\star f_{\rm Hxc}^{\bxi}\right)\star \chi^{\bxi}_{{\rm s},\nu}\star\left(\hat{1}+f_{\rm Hxc}^{\bxi}\star\chi^{\bxi}\right)-\chi^{\bxi}\star\Xi^{\bxi}_{{\rm Hxc},\nu}\star\chi^{\bxi}.
\ee

Returning to Eq.~(\ref{eq:Xinu}), the expression of the effective individual-state Hxc kernel it provides is not the most fruitful one as it neither offers true insight nor permits a practical calculation. We will hence provide suitable alternative expressions to pursue both aims. First, to achieve the form of an actual working equation, we need to remove all reference to any quantity that is not directly available either from the KS ensemble (\ie, the KS orbitals and their energies) or the ensemble Hxc functional (its derivatives with respect to the density and/or the ensemble weights, more specifically). The physical individual density $n_{\Psi_\nu}$, which appears twice in Eq.~(\ref{eq:Xinu}), is problematic in this respect. As Paper I precisely provided a working equation for its evaluation~\cite{Fromager2025indvElevel}, we take advantage of its results to derive in Appendix \ref{sec:appwe} the deviation in density between the true $\nu$th state of interest and its KS analog (or the KS ensemble). This leads to the following explicit formula,
\begin{subequations} \label{eq:weXi}
    \begin{align}
        & \Xi^{\bxi}_{{\rm Hxc},\nu} =  f^{\bxi}_{\text{Hxc}} -\sum_{\lambda>0} (\delta_{\lambda\nu}-\xi_\lambda) \frac{\partial f^{\bxi}_{\text{Hxc}}[n]}{\partial\xi_\lambda}\Bigg|_{n=n^{\bxi}} 
        \\
        & - \left(g^{\bxi}_{\text{Hxc}} \star \left[\big(\hat{1}+\chi^{\bxi} \star  f^{\bxi}_{\text{Hxc}}
     \big) \star \big(n_{\Phi_\nu^{\bxi}}-n^{\bxi}\big) + \chi^{\bxi} \star \sum_{\lambda>0} (\delta_{\lambda\nu}-\xi_\lambda) \frac{\partial v^{\bxi}_{\text{Hxc}}[n]}{\partial\xi_\lambda}\bigg|_{n=n^{\bxi}}\right]\right)
        \\
        & -  \big(\chi^{\bxi}_{\rm s}\big)^{-1} \star \Bigg(\chi^{[2]\bxi}_{\rm s} \star \big(\chi^{\bxi}_{\rm s}\big)^{-1} \star \chi^{\bxi} 
        \\
        &\quad \star\bigg[  f^{\bxi}_{\text{Hxc}} \star \big(n_{\Phi_\nu^{\bxi}}-n^{\bxi}\big) +  \sum_{\lambda>0} (\delta_{\lambda\nu}-\xi_\lambda) \frac{\partial v^{\bxi}_{\text{Hxc}}[n]}{\partial\xi_\lambda}\bigg|_{n=n^{\bxi}}   \bigg]\Bigg) \star \big(\chi^{\bxi}_{\rm s}\big)^{-1}, 
    \end{align}
\end{subequations}
which shows with the ensemble Dyson Eq.~(\ref{eq:ensdyson}) that each effective component of the ensemble Hxc kernel can ultimately be determined, in principle exactly, from the ensemble Hxc functional and its density-functional/weight derivatives.\\    

Now turning to the formal analysis of the individual-state Dyson Eq.~(\ref{eq:dysonlike}), we seek a compact and insightful alternative to the above bulky expression of the individual effective Hxc kernel. Most terms can be gathered into weight derivatives of the ensemble Hxc kernel $f^{\bxi}_{\text{Hxc}}$ and potential $v^{\bxi}_{\text{Hxc}}$ (both being evaluated at the {\it weight-dependent} ensemble density $n^{\bxi}$). For example, from Eq.~\eqref{eq:npsinxi}, one easily sees that
\begin{equation} \label{eq:fdfhxc}
    \sum_{\lambda>0} (\delta_{\lambda\nu}-\xi_\lambda) \frac{\partial f^{\bxi}_{\text{Hxc}}[n]}{\partial\xi_\lambda}\Bigg|_{n=n^{\bxi}}  +  \left(g^{\bxi}_{\text{Hxc}} \star (n_{\Psi_\nu}-n^{\bxi})\right) = \sum_{\lambda>0} (\delta_{\lambda\nu}-\xi_\lambda) \frac{\partial f^{\bxi}_{\text{Hxc}}}{\partial\xi_\lambda}.
\end{equation} 
Inserting Eqs.~(\ref{eq:fdfhxc},\ref{eq:fdvhxc}) into the original form of the individual Hxc kernel component (see Eq.~(\ref{eq:Xinu})) leads to the compact expression,

\begin{equation} \label{eq:Xinu2}
\begin{split}
        \Xi^{\bxi}_{{\rm Hxc},\nu} &= f^{\bxi}_{\text{Hxc}} - \sum_{\lambda>0} (\delta_{\lambda\nu}-\xi_\lambda) \frac{\partial f^{\bxi}_{\text{Hxc}}}{\partial\xi_\lambda} 
        \\
        &\quad
        -  \big(\chi^{\bxi}_{\rm s}\big)^{-1} \star \left(  \chi^{[2]\bxi}_{\rm s} \star \sum_{\lambda>0} (\delta_{\lambda\nu}-\xi_\lambda) \frac{\partial v^{\bxi}_{\text{Hxc}}}{\partial\xi_\lambda} \right) \star \big(\chi^{\bxi}_{\rm s}\big)^{-1}, 
\end{split}
\end{equation}
where we recognize, by analogy with the extraction of individual densities from the ensemble one (see Eq.~(\ref{eq:npsinxi})), effective individual Hxc kernels and potentials. Their physical meaning is further discussed in Sec.~\ref{sec:ind_pot_kernels}. For completeness, we show in Appendix~\ref{sectionapp:extrac_LR_func_from_1st_deriv} that Eq.~(\ref{eq:Xinu2}) is consistent with the linear variation of the exact ensemble linear response function with respect to the ensemble weights. Indeed, the latter property can be exploited to extract directly, through first-order derivatives with respect to the weights, individual-state linear response functions (similarly to Eq.~(\ref{eq:npsinxi}) for the densities).

\section{Discussion}\label{sec:discussion}

We proved in Sec.~\ref{sec:results} that static linear response properties of excited states can be determined, in principle exactly, from the non-interacting ensemble KS system (and proper ensemble density-functional Hxc corrections). In this task, the weight-dependence of the Hxc ensemble density functional and its density-functional derivatives play a crucial role. We explore in the following some fundamental implications of our derivations regarding the construction of Hxc kernels for excited states (Sec.~\ref{sec:ind_pot_kernels}), the connection of the present theory with regular (ground-state) linear response DFT (Sec.~\ref{sec:zero_weight_limit}), and, finally, how instructive the within-the-ensemble ground-state response can be (Sec.~\ref{sec:learning_from_GS_within_ensemble}).

\subsection{Defining individual Hxc kernels for excited states}\label{sec:ind_pot_kernels}

By analogy with the extraction of individual-state properties from the ensemble one (like the density in Eq.~(\ref{eq:npsinxi}), for example), the $\nu$th component $v^{\bxi}_{\text{Hxc},\nu}$ of the ensemble Hxc potential $v^{\bxi}_{\text{Hxc}}$ has been (arbitrarily in fact) defined in Ref.~\citenum{Fromager2025indvElevel} as follows, 
\begin{equation}
   v^{\bxi}_{\text{Hxc},\nu} - v^{\bxi}_{\text{Hxc}}:= \sum_{\lambda>0} (\delta_{\lambda\nu}-\xi_\lambda) \frac{\partial v^{\bxi}_{\text{Hxc}}}{\partial\xi_\lambda}. 
\end{equation}
For now we simply extend the definition to the Hxc kernel (its physical meaning will be discussed later below):
\begin{equation}\label{eq:ind_Hxc_kernel}
    f^{\bxi}_{\text{Hxc},\nu} - f^{\bxi}_{\text{Hxc}}:=\sum_{\lambda>0} (\delta_{\lambda\nu}-\xi_\lambda) \frac{\partial f^{\bxi}_{\text{Hxc}}}{\partial\xi_\lambda}. 
\end{equation}
Note that, by construction (see the ensemble normalization condition in Eq.~(\ref{eq:ens_normalization})),
\be
\begin{split}
\left(\sum_{\nu\geq 0}\xi_\nu v^{\bxi}_{\text{Hxc},\nu}\right)-v^{\bxi}_{\text{Hxc}}
&
=\sum_{\nu\geq 0}\xi_\nu \left(v^{\bxi}_{\text{Hxc},\nu}-v^{\bxi}_{\text{Hxc}}\right)
\\
&=\sum_{\nu\geq 0}\sum_{\lambda>0}\xi_\nu\delta_{\lambda\nu}\frac{\partial v^{\bxi}_{\text{Hxc}}}{\partial\xi_\lambda}-\sum_{\lambda>0}\xi_\lambda\frac{\partial v^{\bxi}_{\text{Hxc}}}{\partial\xi_\lambda} 
\\
&=\sum_{\nu>0}\xi_\nu\frac{\partial v^{\bxi}_{\text{Hxc}}}{\partial\xi_\nu}-\sum_{\lambda>0}\xi_\lambda\frac{\partial v^{\bxi}_{\text{Hxc}}}{\partial\xi_\lambda}
\\
&=0,
\end{split}
\ee
and, similarly,
\be\label{eq:average_ind_kernel_nat_def}
\sum_{\nu\geq 0}\xi_\nu f^{\bxi}_{\text{Hxc},\nu}=f^{\bxi}_{\text{Hxc}}.
\ee
As illustrated by Eq.~(\ref{eq:average_ind_kernel_nat_def}) and Eqs.~(\ref{eq:average_ind_kernels_Xi_def},\ref{eq:Xinu2}),
where the effective individual-state Hxc kernel can be rewritten as follows,
\begin{equation} \label{eq:Xinu3}
        \Xi^{\bxi}_{{\rm Hxc},\nu} = 2 f^{\bxi}_{\text{Hxc}} - f^{\bxi}_{\text{Hxc},\nu}   -  \big(\chi^{\bxi}_{\rm s}\big)^{-1} \star \left(  \chi^{[2]\bxi}_{\rm s} \star \Big(v^{\bxi}_{\text{Hxc},\nu} - v^{\bxi}_{\text{Hxc}}\Big) \right)\star \big(\chi^{\bxi}_{\rm s}\big)^{-1}, 
\end{equation}
there is no unique way to define the individual components $Q^{\bxi}_\nu$ of an ensemble quantity $Q^{\bxi}$, as the latter is invariant under any transformation of type $Q^{\bxi}_\nu\rightarrow Q^{\bxi}_\nu+\sum_{\lambda>0} (\delta_{\lambda\nu}-\xi_\lambda)\Delta^{\bxi}_\lambda$. Therefore, it is not completely clear at this point what the definition of an excited-state Hxc kernel should be in the present (ensemble) context.
Following Ref.~\citenum{Fromager2025indvElevel}, Eq.~\eqref{eq:true_dens_from_stat_cond}, that can now be rewritten as follows,
\begin{equation}
    n_{\Psi_\nu} - n_{\Phi_\nu^{\bxi}}
= \chi_{\rm s}^{\bxi}\star \Big(v^{\bxi}_{\text{Hxc},\nu} - v^{\bxi}_{\text{Hxc}}\Big),
\end{equation}
brings a first enlightening limiting case to mind. Indeed, in the hypothetical (unlikely, in general) situation where the $\nu$th state of the KS ensemble and the true $\nu$th state share the same density, the Hxc potential that reproduces the latter density ($v^{\bxi}_{\text{Hxc}}$ in this special case) would actually be equal to $v^{\bxi}_{\text{Hxc},\nu}$,
\begin{equation}\label{eq:same_dens_same_pot}
    n_{\Psi_\nu} = n_{\Phi_\nu^{\bxi}} \; \Longleftrightarrow \; v^{\bxi}_{\text{Hxc}}=v^{\bxi}_{\text{Hxc},\nu},
\end{equation}
thus giving it a physical meaning, at least from this perspective. An interesting and straightforward implication of Eq.~(\ref{eq:same_dens_same_pot}) is that, according to Eq.~(\ref{eq:Xinu3}),
\begin{equation} \label{eq:Xinu_same_dens}
        n_{\Psi_\nu} = n_{\Phi_\nu^{\bxi}} \; \Longleftrightarrow \;\Xi^{\bxi}_{{\rm Hxc},\nu} = 2 f^{\bxi}_{\text{Hxc}} - f^{\bxi}_{\text{Hxc},\nu},
\end{equation}
which provides a simple (but less intuitive than that of $f^{\bxi}_{\text{Hxc},\nu}$ in Eq.~(\ref{eq:ind_Hxc_kernel})) expression for the effective individual Hxc kernel (from which individual components of the inverse ensemble linear response function are recovered, according to our key Dyson-like Eq.~(\ref{eq:dysonlike})). With the {\it additional} (even more hypothetical) assumption that only the ensemble Hxc kernel is needed to retrieve the exact $\nu$th component of the inverse ensemble linear response function, then the two individual Hxc kernel expressions become equivalent, according to Eq.~(\ref{eq:Xinu_same_dens}): 
\begin{equation}
\label{eq:special_case_all_kernels_equivalent}
\begin{split}
&(R^{\bxi}_{{\rm s},\nu})^{-1}- (R^{\bxi}_{\nu})^{-1}= (\chi^{\bxi}_{\rm s})^{-1}-(\chi^{\bxi})^{-1}
=f^{\bxi}_{\text{Hxc}} 
\\
&
\quad\Longleftrightarrow \; f^{\bxi}_{\text{Hxc}}=\Xi^{\bxi}_{{\rm Hxc},\nu} =f^{\bxi}_{\text{Hxc},\nu},
\end{split}
\end{equation}
thus giving a physical meaning to $f^{\bxi}_{\text{Hxc},\nu}$, at least from this perspective. 

\subsection{Zero-weight limit of the theory and its connection to regular linear response KS-DFT}\label{sec:zero_weight_limit}

Ensemble DFT does not lose rigorous grounds in the limit where all weights $\xi_\lambda$ approach 0. Let us stress that the latter (denoted $\bxi\rightarrow 0^+$) limit of ensemble DFT, from which excited-state properties can still be extracted exactly, differs from regular ground-state DFT, which is recovered when $\bxi=0$. As it should, when considering the $\bxi\rightarrow 0^+$ limit of Eq.~(\ref{eq:Rnu}), the ground-state component $\big(R^{\bxi\rightarrow0^+}_{\nu=0}\big)^{-1}$ of the inverse ensemble linear response function reduces to the ground-state density-density linear response function,
\begin{subequations}
    \begin{align}
        \big(R^{\bxi\rightarrow0^+}_{\nu=0}\big)^{-1} = & (\chi^{\bxi\rightarrow0^+})^{-1} \star \chi_0^{\bxi\rightarrow0^+} \star (\chi^{\bxi\rightarrow0^+})^{-1}   
        \\
        = & (\chi_0)^{-1} \star \chi_0 \star (\chi_0)^{-1}   
        \\
        := & (\chi)^{-1}
        ,
    \end{align}
\end{subequations}
and the effective Hxc kernel in Eq.~\eqref{eq:Xinu2} reduces to the regular (ground-state) Hxc kernel (see also Eq.~(\ref{eq:ind_Hxc_kernel})), 
\begin{equation} \label{eq:Xinu0}
        \Xi^{\bxi\rightarrow0^+}_{{\rm Hxc},\nu=0} =  f^{\bxi\rightarrow0^+}_{\text{Hxc},0}=f^{\bxi\rightarrow0^+}_{\text{Hxc}} =: f_{\text{Hxc}},
\end{equation}
so that the regular ground-state Dyson equation is recovered:
\be\label{eq:regular_GS_Dyson_eq}
(\chi)^{-1}=(\chi_{{\rm s}})^{-1}-f_{\text{Hxc}},
\ee
$\chi_{{\rm s}}:=\chi^{\bxi\rightarrow0^+}_{{\rm s},\nu=0}$ being the ground-state KS linear response function.

\bigskip
When any excited state ($\nu>0$) is considered instead, Eq.~\eqref{eq:Xinu2} very elegantly relates the effective excited-state Hxc kernel to that of regular ground-state linear response DFT $f_{\text{Hxc}}$ through weight-derivatives (taken in the ground-state $\bxi\rightarrow 0^+$ limit) of the ensemble Hxc potential and kernel as follows, 
\begin{equation} \label{eq:Xinu2w0}
\begin{split}
        \Xi_{{\rm Hxc},\nu}:&=\Xi^{\bxi\rightarrow0^+}_{{\rm Hxc},\nu}
        \\
        &\underset{\nu>0}{=} f_{\text{Hxc}} - \frac{\partial f^{\bxi}_{\text{Hxc}}}{\partial\xi_\nu}\bigg|_{\bxi\rightarrow0^+}   -  \big(\chi_{\rm s}\big)^{-1} \star \bigg(  \chi^{[2]}_{\rm s} \star \frac{\partial v^{\bxi}_{\text{Hxc}}}{\partial\xi_\nu}\bigg|_{\bxi\rightarrow0^+} \bigg) \star \big(\chi_{\rm s}\big)^{-1}, 
\end{split}
\end{equation}
where $\chi^{[2]}_{\rm s}$ is the regular (ground-state) KS quadratic response function. As a result, according to the above expression and Eq.~(\ref{eq:true_lrfunct_final_expression_from_ens}) [also taken in the $\bxi\rightarrow 0^+$ limit], it becomes possible to evaluate, in principle exactly, any excited-state linear response function from regular KS-DFT orbitals and energies as follows,
\be\label{eq:EX_lrf_from_GS_orbs_step1}
\chi_\nu=\left(\hat{1}+\chi \star f_{\rm Hxc}\right)\star \chi_{{\rm s},\nu}\star\left(\hat{1}+f_{\rm Hxc}\star\chi\right)-\chi\star\Xi_{{\rm Hxc},\nu}\star\chi
,
\ee
where $\chi_{{\rm s},\nu}:=\chi^{\bxi\rightarrow0^+}_{{\rm s},\nu}$ is the KS linear response function of the ($\nu$th) state under study. Interestingly, by exploiting the ground-state Dyson Eq.~(\ref{eq:regular_GS_Dyson_eq}), we can deduce the following relation, 
\be\label{eq:simp_using_GS_Dyson_eq}
\begin{split}
\chi\star f_{\rm Hxc}\star\chi
&=\chi\star\left(\hat{1}+f_{\rm Hxc}\star\chi\right)-\chi
\\
&=
\left(\hat{1}+\chi \star f_{\rm Hxc}\right)\star \chi_{{\rm s}}\star\left(\hat{1}+f_{\rm Hxc}\star\chi\right)-\chi
,
\end{split}
\ee
which, when combined with Eqs.~(\ref{eq:Xinu2w0}) and (\ref{eq:EX_lrf_from_GS_orbs_step1}), leads to the final compact expression,
\be\label{eq:relation_phys_KS_deviation_EX_from-GS_lrfs}
\chi_\nu-\chi=\left(\hat{1}+\chi \star f_{\rm Hxc}\right)\star \left(\chi_{{\rm s},\nu}-\chi_{{\rm s}}\right)\star\left(\hat{1}+f_{\rm Hxc}\star\chi\right)+\chi\star\tilde{\Xi}_{{\rm Hxc},\nu}\star\chi
,
\ee
where the corrections to the regular ground-state Hxc kernel are now incorporated into 
\be\begin{split}
\tilde{\Xi}_{{\rm Hxc},\nu}&=f_{\rm Hxc}-{\Xi}_{{\rm Hxc},\nu}
\\
&=\frac{\partial f^{\bxi}_{\text{Hxc}}}{\partial\xi_\nu}\bigg|_{\bxi\rightarrow0^+}   +  \big(\chi_{\rm s}\big)^{-1} \star \bigg(  \chi^{[2]}_{\rm s} \star \frac{\partial v^{\bxi}_{\text{Hxc}}}{\partial\xi_\nu}\bigg|_{\bxi\rightarrow0^+} \bigg) \star \big(\chi_{\rm s}\big)^{-1},
\end{split}
\ee
thus establishing an exact relation between the true physical and KS deviations of the excited-state linear response from that of the ground state. Note that Eq.~\eqref{eq:weXi} provides the more explicit working expression,
\begin{subequations} \label{eq:weXi0}
    \begin{align}
        &\tilde{\Xi}_{{\rm Hxc},\nu} =\frac{\partial f^{\bxi}_{\text{Hxc}}[n_0]}{\partial\xi_\nu}\bigg|_{\bxi\rightarrow0^+}
        \\
        & +g_{\text{Hxc}} \star \left[\big(\hat{1}+\chi \star  f_{\text{Hxc}}
     \big) \star \big(n_{\Phi_\nu}-n_0\big) + \chi \star \frac{\partial v^{\bxi}_{\text{Hxc}}[n_0]}{\partial\xi_\nu}\bigg|_{\bxi\rightarrow0^+}\right]
        \\
        & +  \big(\chi_{\rm s}\big)^{-1} \star \left( \chi^{[2]}_{\rm s} \star \big(\chi_{\rm s}\big)^{-1} \star \chi \star\bigg[  f_{\text{Hxc}} \star \big(n_{\Phi_\nu}-n_0\big) + \frac{\partial v^{\bxi}_{\text{Hxc}}[n_0]}{\partial\xi_\nu}\bigg|_{\bxi\rightarrow0^+}   \bigg] \right)\star \big(\chi_{\rm s}\big)^{-1}
        , 
    \end{align}
\end{subequations}
where $n_0=n_{\Phi_0}$ is the exact ground-state density, $n_{\Phi_\nu}$ is the density of the $\nu$th KS state (constructed from regular ground-state KS orbitals), and $g_{\text{Hxc}}\equiv \delta f_{\rm Hxc}/\delta n$ is the regular (ground-state) quadratic Hxc kernel. 
\rev{If, for implementation purposes, we proceed with the following ({\it a priori} drastic) simplification, 
\be\label{eq:simplification_same_Hxc_kernel_for_all_zeroweight_lim}
\tilde{\Xi}_{{\rm Hxc},\nu}\approx 0,
\ee
or, equivalently,  ${\Xi}_{{\rm Hxc},\nu}\approx f_{\rm Hxc}$, Eq.~(\ref{eq:relation_phys_KS_deviation_EX_from-GS_lrfs}) reduces to (according to Eqs.~(\ref{eq:regular_GS_Dyson_eq}) and (\ref{eq:simp_using_GS_Dyson_eq}))
\begin{subequations}
\begin{align}
\label{eq:lrfunc_ES_same_Hxc_kernel_for_all_zeroweight_lim}
\chi_\nu
&\approx \chi\left(\chi_{\rm s}^{-1}\chi_{{\rm s},\nu}\chi_{\rm s}^{-1}-f_{\rm Hxc}\right)\chi
\\
&=\chi\chi_{\rm s}^{-1}\left(\chi_{{\rm s},\nu}-\chi_{\rm s}f_{\rm Hxc}\chi_{\rm s}\right)\chi_{\rm s}^{-1}\chi 
\\
\label{eq:simp_lrfunc_ES_same_Hxc_kernel_for_all_zeroweight_lim}
&\equiv
\chi\chi_{\rm s}^{-1}\left(\hat{1}-\chi_{\rm s}f_{\rm Hxc}\chi_{\rm s}\chi^{-1}_{{\rm s},\nu}\right)\chi_{{\rm s},\nu}\chi_{\rm s}^{-1}\chi
,
\end{align}
\end{subequations}
where we dropped the $\star$ notation, for the sake of compactness. As readily seen from Eq.~(\ref{eq:lrfunc_ES_same_Hxc_kernel_for_all_zeroweight_lim}), this approximation simply consists in making the assumption of Eq.~(\ref{eq:special_case_all_kernels_equivalent}) in the zero-weight limit. A better understanding of its implications, in addition to what has already been discussed in Sec.~\ref{sec:ind_pot_kernels}, is obtained by rewriting Eq.~(\ref{eq:simp_lrfunc_ES_same_Hxc_kernel_for_all_zeroweight_lim}) formally as a Dyson equation,  
\be\label{eq:approx_Dyson_eq_GS_ingredients_only}
\chi^{-1}_\nu&\approx \chi^{-1}_{{\rm s},\nu}-\tilde{f}_{{\rm Hxc},\nu},
\ee
where, according to Eq.~(\ref{eq:regular_GS_Dyson_eq}),
\be
\tilde{f}_{{\rm Hxc},\nu}:=\chi^{-1}_{{\rm s},\nu}-\left(\hat{1}-f_{\rm Hxc}\chi_{\rm s}\right)\chi^{-1}_{{\rm s},\nu}\left(\hat{1}-\chi_{\rm s}f_{\rm Hxc}\chi_{\rm s}\chi^{-1}_{{\rm s},\nu}\right)^{-1}
\left(\hat{1}-\chi_{\rm s}f_{\rm Hxc}\right).
\ee
In order to relate more explicitly the above kernel with that of regular ground-state linear response theory, we use the following equivalent expression,
\be
\begin{split}
\tilde{f}_{{\rm Hxc},\nu}&=\chi^{-1}_{{\rm s},\nu} 
-\Bigg[\left(\hat{1}+f_{\rm Hxc}\left(-\chi_{{\rm s},\nu}+\chi_{{\rm s},\nu}-\chi_{\rm s}\right)\right)\chi^{-1}_{{\rm s},\nu}
\\
&\quad\star\left(\hat{1}-\chi_{\rm s}f_{\rm Hxc}\chi_{\rm s}\chi^{-1}_{{\rm s},\nu}\right)^{-1}
\left(\hat{1}-\chi_{\rm s}f_{\rm Hxc}\left(\chi_{\rm s}+\chi_{{\rm s},\nu}-\chi_{\rm s}\right)\chi^{-1}_{{\rm s},\nu}\right)\Bigg],
\end{split}
\ee
which can then be simplified as follows, 
\be
\begin{split}
\tilde{f}_{{\rm Hxc},\nu}&=\chi^{-1}_{{\rm s},\nu} 
-\Bigg[\left(\chi^{-1}_{{\rm s},\nu}-f_{\rm Hxc}+f_{\rm Hxc}\left(\chi_{{\rm s},\nu}-\chi_{\rm s}\right)\chi^{-1}_{{\rm s},\nu}\right)
\\
&\quad\star
\left(\hat{1}-\left(\hat{1}-\chi_{\rm s}f_{\rm Hxc}\chi_{\rm s}\chi^{-1}_{{\rm s},\nu}\right)^{-1}\chi_{\rm s}f_{\rm Hxc}\left(\chi_{{\rm s},\nu}-\chi_{\rm s}\right)\chi^{-1}_{{\rm s},\nu}\right)\Bigg],
\end{split}
\ee
thus leading to the final expression,
\be\label{eq:final_exp_approx_tilde_kernel_zero_weight_limit}
\tilde{f}_{{\rm Hxc},\nu}=
f_{\rm Hxc}-\left(\hat{1}
-\left(\hat{1}-f_{\rm Hxc}\chi_{\rm s}\right)\chi^{-1}_{{\rm s},\nu}\left(\hat{1}-\chi_{\rm s}f_{\rm Hxc}\chi_{\rm s}\chi^{-1}_{{\rm s},\nu}\right)^{-1}\chi_{\rm s}\right)
f_{\rm Hxc}\left(\chi_{{\rm s},\nu}-\chi_{\rm s}\right)\chi^{-1}_{{\rm s},\nu}.
\ee
As readily seen from the above equation, the (approximate) effective Hxc kernel $\tilde{f}_{{\rm Hxc},\nu}$ for the $\nu$th excited state of interest is determined {\it via} a non-trivial correction to the regular ground-state Hxc kernel $f_{\rm Hxc}$. It uses the exact same ingredients as in a regular linear response KS-DFT calculation, in addition to the linear response function of the $\nu$th excited KS state. Eq.~(\ref{eq:final_exp_approx_tilde_kernel_zero_weight_limit}) also shows that, within the approximation of Eq.~(\ref{eq:simplification_same_Hxc_kernel_for_all_zeroweight_lim}), the difference in Hxc kernel between ground and excited states can be related explicitly to the difference in KS linear response function between the two states. It is unclear at this point how accurate such a scheme would be in practice. A comparison with alternative approaches such as second linear response theory~\cite{SLRTDDFT2016,SLRTDDFT2021}, for example, is expected to bring further insight. This is left for future work.  }\\

\rev{Returning to the exact theory}, as readily seen from Eqs.~(\ref{eq:relation_phys_KS_deviation_EX_from-GS_lrfs}) and (\ref{eq:weXi0}), the weight-dependence of the ensemble density-functional Hxc potential and kernel bring \rev{additional} information about the physical excited states that is not provided by their KS analogs. The weight derivatives are reminiscent of the weight-derivative correction arising in the study of electronic affinity, ionization potential and fundamental gaps through $N$-centered ensemble DFT\cite{senjean2018unified,Cernatic2024_Neutral}. The latter were shown to be equivalent to xc derivative discontinuities present in DFT for fractional electron numbers\cite{perdew1983physical,PRA21_Hodgson_exact_Nc-eDFT_1D}, which are critically missed by practical (semi-)local density-functional approximations, making ensemble DFT a promising alternative as discontinuities are in principle much easier to deal with in practice there\cite{PRA21_Hodgson_exact_Nc-eDFT_1D} (provided, of course, that reliable functionals can be constructed for ensembles~\cite{gould2025ensemblizationdensityfunctionaltheory,Gould2025_PRL_tate-Specific}). The weight derivatives we see in the present work are of a different nature however, being Hxc kernel and potential derivatives rather than Hxc energy derivatives. Ultimately, they relate more closely to the Fukui functions\cite{Hellgren12_Effect,kraisler2013piecewise} but more in-depth analysis of this point within $N$-centered ensemble DFT is left for an upcoming paper. \\

To summarize, the possibility to evaluate excited-state linear response functions {\it directly} on top of a ground-state KS-DFT calculation, to which ensemble corrections \_which are density functionals of the ground-state density\_ are applied, is very appealing from a practical standpoint. Its principle and advantages are reminiscent of what the Direct Ensemble Correction (DEC) scheme achieves to compute excitation energies\cite{yang2017direct,sagredo2018can}, but one should note the development of Paper I together with this work not only yields energies but also densities and now response functions of any excited state. Moreover, the DEC scheme is formally the combination of two ensemble DFT calculations (in the ground-state limit) with a specific weight distribution while we only have to consider one ensemble in our approach, as pointed out in a previous work~\cite{deur2019ground}. Another work which made use of the zero-weight limit of ensemble DFT is the perturbative ensemble generalized KS (pEGKS) approach of Gould and coworkers~\cite{gould2022single,Amoyal_2025_pEDFT}. There, excitation energies can be computed from GKS orbital energy gaps, the zero-weight limit becoming a perturbative limit of the excitation. For completeness, we refer the reader to Refs.~\citenum{Cernatic2024_Neutral,cernatic2024extended_doubles} for an alternative $N$-centered ensemble exactification of KS orbital energy gaps in a more general context.

\subsection{Ground-state properties within the ensemble}\label{sec:learning_from_GS_within_ensemble}

As a final remark and perspective, it is interesting to note that applying Eq.~(\ref{eq:true_lrfunct_final_expression_from_ens}) to the ground state ($\nu=0$) gives the following exact expression for the ground-state linear response function, which holds {for {\it any} ensemble weight values},  
\be\label{eq:GS_lr_func_any_weight}
\chi\underset{\forall \bxi}{=}\left(\hat{1}+\chi^{\bxi}\star f_{\rm Hxc}^{\bxi}\right)\star \chi^{\bxi}_{{\rm s},\nu=0}\star\left(\hat{1}+f_{\rm Hxc}^{\bxi}\star\chi^{\bxi}\right)-\chi^{\bxi}\star\Xi^{\bxi}_{{\rm Hxc},\nu=0}\star\chi^{\bxi},
\ee
where (see Eq.~(\ref{eq:Xinu2}))
\be
\begin{split}
        \Xi^{\bxi}_{{\rm Hxc},\nu=0} &= f^{\bxi}_{\text{Hxc}}+ \sum_{\lambda>0} \xi_\lambda\frac{\partial f^{\bxi}_{\text{Hxc}}}{\partial\xi_\lambda} 
        \\
        &\quad
        +  \big(\chi^{\bxi}_{\rm s}\big)^{-1} \star \left(  \chi^{[2]\bxi}_{\rm s} \star \sum_{\lambda>0} \xi_\lambda\frac{\partial v^{\bxi}_{\text{Hxc}}}{\partial\xi_\lambda} \right) \star \big(\chi^{\bxi}_{\rm s}\big)^{-1}
        .
\end{split}
\ee
The fact that the right-hand side of Eq.~(\ref{eq:GS_lr_func_any_weight}) is {\it weight-independent} is an exact property that could be exploited in the construction of ensemble density-functional approximations, as discussed further below. Since, according to the ensemble Dyson Eq.~(\ref{eq:ensdyson}),
\be
\chi^{\bxi}\star f^{\bxi}_{\text{Hxc}}\star\chi^{\bxi}=
\left(\hat{1}+\chi^{\bxi}\star f_{\rm Hxc}^{\bxi}\right)\star \chi^{\bxi}_{{\rm s}}\star\left(\hat{1}+f_{\rm Hxc}^{\bxi}\star\chi^{\bxi}\right)-\chi^{\bxi},
\ee
this constraint can alternatively be recast into the following exact relation between the true interacting and KS deviations of the ensemble linear response function from the ground-state one:   
\be\label{eq:diff_ens_GS_lrf}
\begin{split}
\chi^{\bxi}-\chi
&\underset{\forall \bxi}{=}
\Bigg[\left(\hat{1}+\chi^{\bxi}\star f_{\rm Hxc}^{\bxi}\right)\star \left(\chi^{\bxi}_{{\rm s}}-\chi^{\bxi}_{{\rm s},\nu=0}
+\left(  \chi^{[2]\bxi}_{\rm s} \star \sum_{\lambda>0} \xi_\lambda\frac{\partial v^{\bxi}_{\text{Hxc}}}{\partial\xi_\lambda} \right)
\right)
\\
&\quad \star\left(\hat{1}+f_{\rm Hxc}^{\bxi}\star\chi^{\bxi}\right)\Bigg]
+\chi^{\bxi}\star \left(\sum_{\lambda>0} \xi_\lambda\frac{\partial f^{\bxi}_{\text{Hxc}}}{\partial\xi_\lambda}\right)\star\chi^{\bxi}
.
\end{split}
\ee
We may use, for example, a simple ansatz that recycles the regular ground-state functional $E_{\rm Hxc}[n]$ as follows, $E^{\bxi}_{\rm Hxc}[n]\approx s(\bxi)E_{\rm Hxc}[n]$, thus ensuring that the exact uniform coordinate scaling constraint, which also holds for ensembles~\cite{Nagy_ensAC,Scott2024_Exact}, is still fulfilled (if, of course, the approximation to $E_{\rm Hxc}[n]$ does fulfill that same constraint, as it should). The weight-dependent scaling function $s(\bxi)$, which equals 1 when $\bxi=0$, by construction (a polynomial function of a given order should be tested first, for convenience and simplicity), would have to be determined from Eqs.~(\ref{eq:ensdyson}) and (\ref{eq:diff_ens_GS_lrf}), at a given level of approximation to $E_{\rm Hxc}[n]$. This is left for future work.

\section{Conclusions}\label{sec:conclusions}

The in-principle exact ensemble DFT of ground and excited energy levels derived in Ref.~\citenum{Fromager2025indvElevel} has been extended to the description of static (density-density) linear responses for each state individually. From the stationarity of energy levels within the density-functional ensemble, we derived an exact and general relation between the individual true physical and KS linear response functions (see Eq.~\eqref{eq:dblsta2}). They do not obey trivial analogs to the ensemble Dyson-like linear response Eq.~(\ref{eq:ensdyson}), but introducing individual components of the inverse ensemble linear response function allowed us to rewrite Eq.~\eqref{eq:dblsta2} in the form of an individual-state Dyson equation (see Eq.~\eqref{eq:dysonlike}) from which the linear response function of the state of interest can be calculated, in principle exactly (see Eq.~(\ref{eq:true_lrfunct_final_expression_from_ens})). From this setting, an effective Hxc kernel naturally emerges for each state within the ensemble. An exact and implementable expression is provided in Eq.~\eqref{eq:weXi}. Its evaluation requires both linear and quadratic responses of the KS ensemble, together with the physical ensemble linear response (which is deduced from the KS one through the ensemble Dyson Eq.~\eqref{eq:ensdyson}). Not only the density-functional derivatives of the ensemble Hxc energy but also their ensemble weight derivatives play an additional key role in this task.\\  

Various fundamental implications of our derivations have been discussed, starting with the identification of two different definitions for the Hxc kernel (for each state within the ensemble) and their connection under hypothetical mapping assumptions (see Eqs.~\eqref{eq:Xinu2}, \eqref{eq:ind_Hxc_kernel}, and \eqref{eq:special_case_all_kernels_equivalent}). Furthermore, we considered the zero-weight limit of the theory, which infinitesimally deviates from regular ground-state linear response DFT, thus allowing for an in-principle exact description of excited-state linear response properties from the standard KS orbitals and their energies. An explicit connection between the effective excited-state Hxc kernel and the regular (ground-state) one has been established in this limit (see Eq.~(\ref{eq:Xinu2w0})), as well as the compact working Eq.~(\ref{eq:relation_phys_KS_deviation_EX_from-GS_lrfs}) from which the linear response of a specific excited state can be retrieved. As shown explicitly in Eq.~(\ref{eq:weXi0}), the corrections to regular (ground-state) static linear response DFT that are needed to make the zero-weight limit exact involve the first-order derivatives in weights of both the ensemble Hxc potential and kernel. This ground-state KS-DFT + ensemble corrections setting is promising from a computational standpoint as practical ensemble DFT schemes yielding electronic excitations enabled by this very principle have recently emerged, {\ie} DEC with exact ensemble exchange~\cite{yang2017direct,sagredo2018can} (EEXX) or the perturbative ensemble generalized DFT approach~\cite{gould2022single,Amoyal_2025_pEDFT} (pEGDFT) with regular (ground-state) DFAs. Their good performance and ability to tackle situations where LR-TDDFT fails, \rev{namely states with double excitation character and degeneracies,} together with continued development of weight-dependent ensemble DFAs~\cite{Gould2025_PRL_tate-Specific} make ensemble DFT a promising approach, which our in-principle exact and general linear response framework paves the way towards. 
\rev{One may be concerned with a perceived increase in complexity in our approach, as it includes quadratic response functions, but note only those of the KS system appear in our expression, and thus obtaining them is inexpensive and straightforward. Moreover, one might also think we need to solve additional Dyson equations (one for the ensemble Eq.~\eqref{eq:ensdyson}, one for each individual state Eq.~\eqref{eq:dysonlike}) which would constitute a drawback. But it is not the case. Looking at Eq.~\eqref{eq:true_lrfunct_final_expression_from_ens}, the right hand side can be computed without having to know anything more about the true (many-body) system than its ensemble density and density-density response.}
Obviously, extending this work to the time-dependent linear response regime will give access to more properties, such as transition dipole moments between excited states, for example. 
\rev{Upon revision of the present work, we became aware of a preprint~\cite{daas2025ensembletimedependentdensityfunctional} that appeared after our first submission and deals precisely with the time-dependent linear response extension of EDFT, thus giving access to transition frequencies and oscillator strengths, in principle exactly. As it focuses on the response of the ensemble, it does not constitute an extension of our own work. It is nevertheless very promising for the future as limitations of EDFT are being pushed back significantly. Exploration of the time-dependent analog to our formalism is left for future work.}
Finally, by considering for any ensemble weight values the expression of the physical (weight-independent) ground-state linear response function, we derived an exact constraint (see Eq.~(\ref{eq:diff_ens_GS_lrf})) that we expect to be useful for designing ensemble DFAs that recycle ground-state ones. This is also left for future work.

\begin{acknowledgement}
This work has benefited from support provided by the University of Strasbourg Institute for Advanced Study (USIAS) for a Fellowship, within the French national programme “Investment for the future” (IdEx-Unistra).

\end{acknowledgement}
\appendix

\section*{Appendix}

\section{Detailed differentiation of stationarity conditions for individual energy levels}\label{secapp:deriv_stat_cond}

Taking the second functional derivative of Eq.~\eqref{eq:dblsta1} involves liberal use of chain rule, and we'll make the crucial steps more explicit. First, note that
\begin{subequations}\label{eq:trick_func_deriv_inv_KS_ens_rsp_func}
\begin{align}
    {\delta \big(\chi^{\bxi}_{\rm s}\big[n^{\bxi}_v\big] \big)^{-1}}\bigg|_{v=v_{\text{ext}}} & = - \big(\chi^{\bxi}_{\rm s} \big)^{-1} \star {\delta \chi^{\bxi}_{\rm s} \big[n^{\bxi}_v\big]}\bigg|_{v=v_{\text{ext}}} \star \big(\chi^{\bxi}_{\rm s} \big)^{-1}
    \\
    & = - \big(\chi^{\bxi}_{\rm s} \big)^{-1} \star \left(\chi^{[2]\bxi}_{\rm s} \star \left.\delta v_{\rm s}^{\bxi}\big[n^{\bxi}_v\big]\right|_{v=v_{\text{ext}}} \right)\star \big(\chi^{\bxi}_{\rm s}\big)^{-1}
\end{align}
\end{subequations}
where we introduced the quadratic response function of the KS system whose sum-over-states expression is given in Appendix~\ref{app:SOS_KS_quadratic_rsp_func}, 
\be
\chi_{\rm s}^{[2]\bxi}(\br,\br',\br'')\equiv \dfrac{\delta \chi_{\rm s}^{\bxi}(\br,\br')}{\delta v_{\rm s}^{\bxi}(\br'')}
\ee

and, according to the ensemble Dyson Eq.~(\ref{eq:ensdyson}) and Eq.~(\ref{eq:KS_pot_for_any_local_v}),
\begin{equation}\label{eq:delta_KS_pot_func_n_v}
\begin{split}
\left.\delta v_{\rm s}^{\bxi}\big[n^{\bxi}_v\big]\right|_{v=v_{\text{ext}}}
&=\delta v+f^{\bxi}_{\rm Hxc}\star\chi^{\bxi}\star\delta v
\\
&=\left(\hat{1}+f^{\bxi}_{\rm Hxc}\star\chi^{\bxi}\right)\star\delta v
\\
&=\big(\chi^{\bxi}_{\rm s} \big)^{-1} \star \chi^{\bxi}\star\delta v
\end{split}
\end{equation}
Owing to the 3-point dependence of the quadratic response, we need to disambiguate how the $\star$ product applies in Eq.~\eqref{eq:trick_func_deriv_inv_KS_ens_rsp_func}. Our convention is that $\chi_{\rm s}^{[2]\bxi}\star f$ always contracts the last coordinate of $\chi_{\rm s}^{[2]\bxi}$ (see their meaning above) with the first coordinate of $f$. Moreover, the operation order dictated by parentheses is to be respected as in general $(f\star g)\star h \neq f\star (g\star h)$. More explicitly,
\begin{equation}
   \left[f_a \star \big(\chi^{[2]\bxi}_{\rm s} \star n \big) \star f_b \right](\mathbf{r},\mathbf{r}'''') = \int d\mathbf{r}'d\mathbf{r}''d\mathbf{r}''' f_a(\mathbf{r},\mathbf{r}') \; \chi^{[2]\bxi}_{\rm s}(\mathbf{r}',\mathbf{r}'',\mathbf{r}''') \; n(\mathbf{r}''') \; f_b(\mathbf{r}'',\mathbf{r}'''')
\end{equation}
Following this convention, we can write for any density $n$
\begin{equation}
\begin{split}
&\left.\delta \left(\big(\chi^{\bxi}_{\rm s}\big[n^{\bxi}_v\big] \big)^{-1}\star n\right)\right|_{v=v_{\text{ext}}}
\\
&\quad=
- \big(\chi^{\bxi}_{\rm s} \big)^{-1} \star \left(\chi^{[2]\bxi}_{\rm s} \star \big(\chi^{\bxi}_{\rm s} \big)^{-1} \star \chi^{\bxi}\star\delta v \right)\star \big(\chi^{\bxi}_{\rm s}\big)^{-1}\star n
\\
&
\quad
=
- \big(\chi^{\bxi}_{\rm s} \big)^{-1} \star \left(\chi^{[2]\bxi}_{\rm s} \star \big(\chi^{\bxi}_{\rm s} \big)^{-1} \star n\right)\star \big(\chi^{\bxi}_{\rm s}\big)^{-1}\star\chi^{\bxi}\star\delta v
\end{split}
\end{equation}

where we used the following symmetry property
\be
\chi^{[2]\bxi}_{\rm s}(\br,\br',\br'')=\dfrac{\delta^2 n_{\rm s}^{\bxi}(\br)}{\delta v_{\rm s}^{\bxi}(\br')\delta v_{\rm s}^{\bxi}(\br'')}\equiv \dfrac{\delta^2 n_{\rm s}^{\bxi}(\br)}{\delta v_{\rm s}^{\bxi}(\br'')\delta v_{\rm s}^{\bxi}(\br')}=\chi^{[2]\bxi}_{\rm s}(\br,\br'',\br'), 
\ee
thus leading to the compact expression
\be\label{eq:trick_func_deriv_inv_KS_ens_rsp_func_times_n}
 \left.\frac{\delta \left(\big(\chi^{\bxi}_{\rm s}\big[n^{\bxi}_v\big] \big)^{-1}\star n\right)}{\delta v}\right|_{v=v_{\text{ext}}} 
    & =
    - \big(\chi^{\bxi}_{\rm s} \big)^{-1} \star \left(\chi^{[2]\bxi}_{\rm s} \star \big(\chi^{\bxi}_{\rm s} \big)^{-1} \star n\right)\star \big(\chi^{\bxi}_{\rm s}\big)^{-1}\star\chi^{\bxi}
    \ee
    
Moreover, we define
\begin{equation}
    g^{\bxi}_{\text{Hxc}} = g^{\bxi}_{\text{Hxc}}[n^{\bxi}] = \left.\frac{\delta f^{\bxi}_{\text{Hxc}}[n]}{\delta n}\right|_{n=n^{\bxi}}
    \equiv \left.\dfrac{\delta^2 v^{\bxi}_{\rm Hxc}[n](\br)}{\delta n(\br')\delta n(\br'')}\right|_{n=n^{\bxi}}
\end{equation}
with $n^{\bxi}[v_{\text{ext}}]=n^{\bxi}$ so that for any density $n$
\begin{equation}
\begin{split}
\left.\delta\left(f^{\bxi}_{\rm Hxc}\left[n_v^{\bxi}\right]\star n\right)\right|_{v=v_{\text{ext}}}
&=\left(g^{\bxi}_{\text{Hxc}}\star\chi^{\bxi}\star \delta v\right)\star n 
\\
&=\left(g^{\bxi}_{\text{Hxc}}\star n \right)\star \chi^{\bxi}\star \delta v
\end{split}
\end{equation}
making use of the symmetry property $g^{\bxi}_{\text{Hxc}}(\br,\br',\br'')=g^{\bxi}_{\text{Hxc}}(\br,\br'',\br')$. In the end we can write
\begin{equation}
    \left.\frac{\delta}{\delta v}\Bigg[f^{\bxi}_{\text{Hxc}}\big[n^{\bxi}_v\big]\star n\Bigg]\right|_{v=v_{\text{ext}}} = \left(g^{\bxi}_{\text{Hxc}} \star n\right)\star \chi^{\bxi}
\end{equation}
Lastly, we exploit the fact that $n$ and $\xi$ are independent variables to swap derivatives:
\begin{subequations}
\begin{align}
       \left.\frac{\delta}{\delta v}\Bigg[\frac{\partial v^{\bxi}_{\text{Hxc}}[n]}{\partial\xi_\lambda}\Bigg|_{n=n^{\bxi}_v}\Bigg]\right|_{v=v_{\text{ext}}} 
       = & \Bigg[ \frac{\delta}{\delta n}\bigg[\frac{\partial v^{\bxi}_{\text{Hxc}}[n]}{\partial\xi_\lambda}\bigg]\bigg|_{n=n^{\bxi}} \Bigg] \star \chi^{\bxi} 
       \\
       = & \Bigg[ \frac{\partial}{\partial\xi_\lambda} \bigg[\frac{\delta v^{\bxi}_{\text{Hxc}}[n]}{\delta n}\bigg]\bigg|_{n=n^{\bxi}} \Bigg] \star \chi^{\bxi}
       \\
       = & \frac{\partial f^{\bxi}_{\text{Hxc}}[n]}{\partial\xi_\lambda}\Bigg|_{n=n^{\bxi}} \star \chi^{\bxi} 
\end{align}
\end{subequations}
We now follow through with the second differentiation in Eq.~(\ref{eq:dblsta1}) where the individual (ground or excited-state) KS linear response function emerges as follows (see Eq.~(\ref{eq:delta_KS_pot_func_n_v})) 
\be
\left.\dfrac{\delta n_{\Phi_\nu^{\bxi}}\left[n^{\bxi}_v\right]}{\delta v}\right|_{v=v_{\text{ext}}}=\chi^{\bxi}_{{\rm s},\nu}\star\left.\dfrac{\delta v_{\rm s}^{\bxi}\big[n^{\bxi}_v\big]}{\delta v}\right|_{v=v_{\text{ext}}}=\chi^{\bxi}_{{\rm s},\nu}\star \big(\chi^{\bxi}_{\rm s} \big)^{-1} \star \chi^{\bxi}.
\ee
and get an expression relating the individual response function of the many-body system to that of the KS one given in Eq.~(\ref{eq:dblsta2}).

\section{Working equations for the effective components of the ensemble Hxc kernel}\label{sec:appwe}

To derive Eq.~(\ref{eq:weXi}) in main text, we start by combining Eq.~(73a) of Paper I,
\begin{equation}
     \sum_{\lambda>0} (\delta_{\lambda\nu}-\xi_\lambda) \left( \frac{\partial n^{\bxi}}{\partial\xi_\lambda} - 
     \big(n_{\Phi_\lambda^{\bxi}}-n_{\Phi_0^{\bxi}}
     \big) \right)= n_{\Psi_\nu}-n_{\Phi_\nu^{\bxi}}
\end{equation}
with Eq.~(74)
\begin{equation}
     \frac{\partial n^{\bxi}}{\partial\xi_\lambda} - 
     \big(n_{\Phi_\lambda^{\bxi}}-n_{\Phi_0^{\bxi}}
     \big) = \chi^{\bxi}_{\rm s} \star \frac{\partial v^{\bxi}_{\text{Hxc}}}{\partial\xi_\lambda} 
\end{equation}
so that it comes
\begin{equation} \label{eq:fdvhxc}
     n_{\Psi_\nu}-n_{\Phi_\nu^{\bxi}} = \chi^{\bxi}_{\rm s} \star \sum_{\lambda>0} (\delta_{\lambda\nu}-\xi_\lambda) \frac{\partial v^{\bxi}_{\text{Hxc}}}{\partial\xi_\lambda}
\end{equation}
Furthermore, combining Eqs. (\ref{eq:dnxi},\ref{eq:gathering_KS_terms_deriv_ens_dens_wrt_weight}) yields
\begin{equation} \label{eq:ldnxi}
    \sum_{\lambda>0} (\delta_{\lambda\nu}-\xi_\lambda) \frac{\partial n^{\bxi}}{\partial\xi_\lambda} = (\hat{1}+\chi^{\bxi} \star f^{\bxi}_{\text{Hxc}} ) \star (n_{\Phi_\nu^{\bxi}}-n^{\bxi}) + \chi^{\bxi} \star \sum_{\lambda>0} (\delta_{\lambda\nu}-\xi_\lambda) \frac{\partial v^{\bxi}_{\text{Hxc}}}{\partial\xi_\lambda} 
\end{equation}
which we insert in Eq.~(\ref{eq:npsinxi}) to obtain
\begin{equation} \label{eq:diffn}
     n_{\Psi_\nu}-n^{\bxi} = (\hat{1}+\chi^{\bxi} \star f^{\bxi}_{\text{Hxc}} ) \star (n_{\Phi_\nu^{\bxi}}-n^{\bxi}) + \chi^{\bxi} \star \sum_{\lambda>0} (\delta_{\lambda\nu}-\xi_\lambda) \frac{\partial v^{\bxi}_{\text{Hxc}}}{\partial\xi_\lambda} 
\end{equation}
We see that both density differences involve the weight derivative of $v^{\bxi}_{\text{Hxc}}$, which needs to be split in two contributions to be computable:
\begin{equation} \label{eq:dvhxcf}
    \frac{\partial v^{\bxi}_{\text{Hxc}}}{\partial\xi_\lambda} = \frac{\partial v^{\bxi}_{\text{Hxc}}[n]}{\partial\xi_\lambda}\bigg|_{n=n^{\bxi}}  + f^{\bxi}_{\text{Hxc}} \star \frac{\partial n^{\bxi}}{\partial\xi_\lambda}
\end{equation}
We combine Eq.~(\ref{eq:ldnxi}) with Eq.~\eqref{eq:dvhxcf} to write:
\begin{equation} \label{eq:fdvhxcfinal}
    \sum_{\lambda>0} (\delta_{\lambda\nu}-\xi_\lambda) \frac{\partial v^{\bxi}_{\text{Hxc}}}{\partial\xi_\lambda} = (\chi^{\bxi}_{\rm s})^{-1} \star \chi^{\bxi} \left(  \sum_{\lambda>0} (\delta_{\lambda\nu}-\xi_\lambda) \frac{\partial v^{\bxi}_{\text{Hxc}}[n]}{\partial\xi_\lambda}\bigg|_{n=n^{\bxi}} + f^{\bxi}_{\text{Hxc}} \star \big(n_{\Phi_\nu^{\bxi}}-n^{\bxi}\big) \right)
\end{equation}
Finally, Eq.~(\ref{eq:fdvhxcfinal}) can be inserted in Eq.~(\ref{eq:fdvhxc}) we obtain:
\begin{equation} \label{eq:diffpsiphi}
    n_{\Psi_\nu}-n_{\Phi_\nu^{\bxi}} = \chi^{\bxi} \star \left(  \sum_{\lambda>0} (\delta_{\lambda\nu}-\xi_\lambda) \frac{\partial v^{\bxi}_{\text{Hxc}}[n]}{\partial\xi_\lambda}\bigg|_{n=n^{\bxi}} + f^{\bxi}_{\text{Hxc}} \star \big(n_{\Phi_\nu^{\bxi}}-n^{\bxi}\big) \right)
\end{equation}
while inserting Eq.~(\ref{eq:fdvhxcfinal}) in Eq.~(\ref{eq:diffn}) yields
\begin{equation} \label{eq:diffpsifull}
     n_{\Psi_\nu}-n^{\bxi} = 
     \big(1+\chi^{\bxi} \star  f^{\bxi}_{\text{Hxc}}
     \big) \star \big(n_{\Phi_\nu^{\bxi}}-n^{\bxi}\big) + \chi^{\bxi} \star \sum_{\lambda>0} (\delta_{\lambda\nu}-\xi_\lambda) \frac{\partial v^{\bxi}_{\text{Hxc}}[n]}{\partial\xi_\lambda}\bigg|_{n=n^{\bxi}}  
\end{equation}
Combining Eqs.~(\ref{eq:diffpsifull},\ref{eq:diffpsiphi},\ref{eq:Xinu}) allows to retrieve Eq.~(\ref{eq:weXi}) in main text.

\section{Sum-over-states expression of the ensemble KS quadratic response function}\label{app:SOS_KS_quadratic_rsp_func}
The quadratic response function of the KS ensemble is evaluated as follows,      
\be
\chi_{\rm s}^{[2]\bxi}(\br,\br',\br'')\equiv \sum_{\kappa\geq 0}\xi_\kappa\dfrac{\delta \chi^{\bxi}_{{\rm
s},\kappa}(\br,\br')}{\delta v_{\rm s}^{\bxi}(\br'')}=\sum_{\kappa\geq 0}\xi_\kappa \chi_{{\rm
s},\kappa}^{[2]\bxi}(\br,\br',\br''),
\ee
where individual-state quadratic response functions are obtained by differentiating the linear response functions
\bse
\begin{align}
\label{eq:ind_KS_state_dens-dens_rsp_func}
\chi^{\bxi}_{{\rm
s},\kappa}({\br},{\br'})
&=\dfrac{\delta n_{\Phi^{\bxi}_\kappa}(\br)}{\delta v^{\bxi}_{\rm s}(\br')}
\\
\label{eq:ind_KS_state_dens-dens_rsp_func_part2}
&=2\sum_{\mu\neq
\kappa}\dfrac{\langle\Phi^{\bxi}_\kappa\vert\hat{n}({\br})\vert
\Phi^{\bxi}_\mu\rangle\langle\Phi^{\bxi}_\mu\vert\hat{n}({\br}')\vert
\Phi^{\bxi}_\kappa\rangle}{\mathcal{E}^{\bxi}_{\kappa}-\mathcal{E}^{\bxi}_{\mu}}
\end{align}
\ese
with respect to the ensemble KS potential. Since we have through first order in $\delta v^{\bxi}_{\rm s}$, 
\bse
\begin{align}
\dfrac{1}{2}\delta\chi^{\bxi}_{{\rm
s},\kappa}({\br},{\br'})
&=
\dfrac{1}{2}\left(\chi^{\bxi}_{{\rm
s},\kappa}[v_{\rm s}^{\bxi}+\delta v_{\rm s}^{\bxi}]({\br},{\br'})-\chi^{\bxi}_{{\rm
s},\kappa}[v_{\rm s}^{\bxi}]({\br},{\br'})\right) 
\\
&=\sum_{\mu\neq
\kappa}\dfrac{\langle\delta\Phi^{\bxi}_\kappa\vert\hat{n}({\br})\vert
\Phi^{\bxi}_\mu\rangle\langle\Phi^{\bxi}_\mu\vert\hat{n}({\br}')\vert
\Phi^{\bxi}_\kappa\rangle}{\mathcal{E}^{\bxi}_{\kappa}-\mathcal{E}^{\bxi}_{\mu}}
\\
&\quad+\sum_{\mu\neq
\kappa}\dfrac{\langle\Phi^{\bxi}_\kappa\vert\hat{n}({\br})\vert
\delta\Phi^{\bxi}_\mu\rangle\langle\Phi^{\bxi}_\mu\vert\hat{n}({\br}')\vert
\Phi^{\bxi}_\kappa\rangle}{\mathcal{E}^{\bxi}_{\kappa}-\mathcal{E}^{\bxi}_{\mu}}
\\
&\quad+\sum_{\mu\neq
\kappa}\dfrac{\langle\Phi^{\bxi}_\kappa\vert\hat{n}({\br})\vert
\Phi^{\bxi}_\mu\rangle\langle\delta\Phi^{\bxi}_\mu\vert\hat{n}({\br}')\vert
\Phi^{\bxi}_\kappa\rangle}{\mathcal{E}^{\bxi}_{\kappa}-\mathcal{E}^{\bxi}_{\mu}}
\\
&\quad+\sum_{\mu\neq
\kappa}\dfrac{\langle\Phi^{\bxi}_\kappa\vert\hat{n}({\br})\vert
\Phi^{\bxi}_\mu\rangle\langle\Phi^{\bxi}_\mu\vert\hat{n}({\br}')\vert \delta
\Phi^{\bxi}_\kappa\rangle}{\mathcal{E}^{\bxi}_{\kappa}-\mathcal{E}^{\bxi}_{\mu}}
\\
&\quad-\sum_{\mu\neq
\kappa}\dfrac{\left(\delta\mathcal{E}^{\bxi}_{\kappa}-\delta\mathcal{E}^{\bxi}_{\mu}\right)\langle\Phi^{\bxi}_\kappa\vert\hat{n}({\br})\vert
\Phi^{\bxi}_\mu\rangle\langle\Phi^{\bxi}_\mu\vert\hat{n}({\br}')\vert
\Phi^{\bxi}_\kappa\rangle}{\left(\mathcal{E}^{\bxi}_{\kappa}-\mathcal{E}^{\bxi}_{\mu}\right)^2}
,
\end{align}
\ese
where
\bse
\begin{align}
\myket{\delta\Phi^{\bxi}_\kappa}&=\sum_{ \tau\neq\kappa}\myket{\Phi^{\bxi}_\tau}
\int d\br''\;\dfrac{\langle\Phi^{\bxi}_\tau\vert\hat{n}({\br''})\vert
\Phi^{\bxi}_\kappa\rangle}{\mathcal{E}^{\bxi}_{\kappa}-\mathcal{E}^{\bxi}_{\tau}}\delta v_{\rm s}^{\bxi}({\br''}),
\\
\myket{\delta\Phi^{\bxi}_\mu}&=\sum_{ \mu'\neq\mu}\myket{\Phi^{\bxi}_{\mu'}}
\int d\br''\;\dfrac{\langle\Phi^{\bxi}_{\mu'}\vert\hat{n}({\br''})\vert
\Phi^{\bxi}_\mu\rangle}{\mathcal{E}^{\bxi}_{\mu}-\mathcal{E}^{\bxi}_{\mu'}}\delta v_{\rm s}^{\bxi}({\br''}),
\end{align}
\ese
and
\bse
\begin{align}
\delta\mathcal{E}^{\bxi}_{\kappa}=\int d\br''\;\langle\Phi^{\bxi}_{\kappa}\vert\hat{n}({\br''})\vert
\Phi^{\bxi}_\kappa\rangle\delta v_{\rm s}^{\bxi}({\br''}),
\\
\delta\mathcal{E}^{\bxi}_{\mu}=\int d\br''\;\langle\Phi^{\bxi}_{\mu}\vert\hat{n}({\br''})\vert
\Phi^{\bxi}_\mu\rangle\delta v_{\rm s}^{\bxi}({\br''}),
\end{align}
\ese
thus leading to
\bse
\begin{align}
\dfrac{1}{2}\chi_{{\rm
s},\kappa}^{[2]\bxi}(\br,\br',\br'')
&=\sum_{\mu\neq
\kappa}
\sum_{\tau\neq\kappa}
\dfrac{\langle\Phi^{\bxi}_\tau\vert\hat{n}({\br})\vert
\Phi^{\bxi}_\mu\rangle\langle\Phi^{\bxi}_\mu\vert\hat{n}({\br}')\vert
\Phi^{\bxi}_\kappa\rangle\langle\Phi^{\bxi}_\kappa\vert\hat{n}({\br''})\vert
\Phi^{\bxi}_\tau\rangle}{\left(\mathcal{E}^{\bxi}_{\kappa}-\mathcal{E}^{\bxi}_{\tau}\right)\left(\mathcal{E}^{\bxi}_{\kappa}-\mathcal{E}^{\bxi}_{\mu}\right)}
\\
&\quad+\sum_{\mu\neq
\kappa}\sum_{ \mu'\neq\mu}\dfrac{\langle\Phi^{\bxi}_\kappa\vert\hat{n}({\br})\vert
\Phi^{\bxi}_{\mu'}\rangle\langle\Phi^{\bxi}_\mu\vert\hat{n}({\br}')\vert
\Phi^{\bxi}_\kappa\rangle\langle\Phi^{\bxi}_{\mu'}\vert\hat{n}({\br''})\vert
\Phi^{\bxi}_\mu\rangle}{\left(\mathcal{E}^{\bxi}_{\mu}-\mathcal{E}^{\bxi}_{\mu'}\right)\left(\mathcal{E}^{\bxi}_{\kappa}-\mathcal{E}^{\bxi}_{\mu}\right)}
\\
&\quad+\sum_{\mu\neq
\kappa}
\sum_{ \mu'\neq\mu}
\dfrac{\langle\Phi^{\bxi}_\kappa\vert\hat{n}({\br})\vert
\Phi^{\bxi}_\mu\rangle\langle\Phi^{\bxi}_{\mu'}\vert\hat{n}({\br}')\vert
\Phi^{\bxi}_\kappa\rangle\langle\Phi^{\bxi}_{\mu}\vert\hat{n}({\br''})\vert
\Phi^{\bxi}_{\mu'}\rangle}{\left(\mathcal{E}^{\bxi}_{\mu}-\mathcal{E}^{\bxi}_{\mu'}\right)\left(\mathcal{E}^{\bxi}_{\kappa}-\mathcal{E}^{\bxi}_{\mu}\right)}
\\
&\quad+\sum_{\mu\neq
\kappa}\sum_{\tau\neq\kappa}\dfrac{\langle\Phi^{\bxi}_\kappa\vert\hat{n}({\br})\vert
\Phi^{\bxi}_\mu\rangle\langle\Phi^{\bxi}_\mu\vert\hat{n}({\br}')\vert 
\Phi^{\bxi}_\tau\rangle\langle\Phi^{\bxi}_\tau\vert\hat{n}({\br''})\vert
\Phi^{\bxi}_\kappa\rangle}{\left(\mathcal{E}^{\bxi}_{\kappa}-\mathcal{E}^{\bxi}_{\tau}\right)\left(\mathcal{E}^{\bxi}_{\kappa}-\mathcal{E}^{\bxi}_{\mu}\right)}
\\
&\quad-\sum_{\mu\neq
\kappa}\dfrac{\langle\Phi^{\bxi}_\kappa\vert\hat{n}({\br})\vert
\Phi^{\bxi}_\mu\rangle\langle\Phi^{\bxi}_\mu\vert\hat{n}({\br}')\vert
\Phi^{\bxi}_\kappa\rangle
\left(n_{\Phi^{\bxi}_{\kappa}}(\br'')-n_{\Phi^{\bxi}_{\mu}}(\br'')\right)
}{\left(\mathcal{E}^{\bxi}_{\kappa}-\mathcal{E}^{\bxi}_{\mu}\right)^2}.
\end{align}
\ese


\section{Alternative derivation of the Dyson equation for individual components of the inverse ensemble linear response function}\label{sectionapp:extrac_LR_func_from_1st_deriv}

Like the true physical ensemble energy and density, the exact ensemble density-density linear response function varies linearly with the ensemble weights,
\be\label{eqapp:true_ens_rsp_func_exp}
\chi^{\bxi}=\sum_{\nu\geq 0}\xi_\nu\chi_\nu=\chi_0+\sum_{\lambda>0}\xi_\lambda\left(\chi_\lambda-\chi_0\right),
\ee
so that individual-state response functions can be extracted, in principle exactly, as follows, 
\be
\chi_\nu\underset{\nu\geq 0}{=}\chi^{\bxi}+\sum_{\lambda>0}\left(\delta_{\lambda\nu}-\xi_\lambda\right)\dfrac{\partial \chi^{\bxi}}{\partial \xi_\lambda},
\ee
thus leading to (see Eq.~(\ref{eq:Rnu}))
\be\label{eqapp:invRnu_from_linearity}
(R^{\bxi}_\nu)^{-1}=\left(\chi^{\bxi}\right)^{-1}+\sum_{\lambda>0}\left(\delta_{\lambda\nu}-\xi_\lambda\right)\left(\chi^{\bxi}\right)^{-1}\dfrac{\partial \chi^{\bxi}}{\partial \xi_\lambda}\left(\chi^{\bxi}\right)^{-1}. 
\ee
Since, by analogy with Eq.~(\ref{eq:trick_func_deriv_inv_KS_ens_rsp_func}), 
\be
\left(\chi^{\bxi}\right)^{-1}\dfrac{\partial \chi^{\bxi}}{\partial \xi_\lambda}+\left(\dfrac{\partial \left(\chi^{\bxi}\right)^{-1}}{\partial \xi_\lambda}\right)\chi^{\bxi}=\dfrac{\partial}{\partial\xi_\lambda}\left(\hat{1}\right)=0,
\ee
or, equivalently,
\be
\left(\chi^{\bxi}\right)^{-1}\dfrac{\partial \chi^{\bxi}}{\partial \xi_\lambda}\left(\chi^{\bxi}\right)^{-1}=-\dfrac{\partial \left(\chi^{\bxi}\right)^{-1}}{\partial \xi_\lambda},
\ee
it comes from Eq.~(\ref{eqapp:invRnu_from_linearity}) and the ensemble Dyson Eq.~(\ref{eq:ensdyson}),
\be\label{eq:inv_rsp_compt_starting_eq_from_linearity}
(R^{\bxi}_\nu)^{-1}=\left(\chi_{\rm s}^{\bxi}\right)^{-1}-f_{\rm Hxc}^{\bxi}-\sum_{\lambda>0}\left(\delta_{\lambda\nu}-\xi_\lambda\right)\left(\dfrac{\partial \left(\chi_{\rm s}^{\bxi}\right)^{-1}}{\partial \xi_\lambda}-\dfrac{\partial f_{\rm Hxc}^{\bxi}}{\partial \xi_\lambda}\right).
\ee
Similarly, we have for the KS ensemble, 
\be\label{eq:deriv_inv_ens_rsp_KS}
\dfrac{\partial \left(\chi_{\rm s}^{\bxi}\right)^{-1}}{\partial \xi_\lambda}=-\left(\chi_{\rm s}^{\bxi}\right)^{-1}\dfrac{\partial \chi_{\rm s}^{\bxi}}{\partial \xi_\lambda}\left(\chi_{\rm s}^{\bxi}\right)^{-1},
\ee
where
\be
\dfrac{\partial \chi_{\rm s}^{\bxi}}{\partial \xi_\lambda}\equiv \dfrac{\partial }{\partial \xi_\lambda}\left(\chi_{\rm s}^{\bxi}\left[v_{\rm s}^{\bxi}\right]\right)=\chi^{\bxi}_{{\rm s},\lambda}-\chi^{\bxi}_{{\rm s},0}+\left.\dfrac{\partial \chi_{\rm s}^{\bm \zeta}\left[v_{\rm s}^{\bxi}\right]}{\partial \xi_\lambda}\right|_{{\bm \zeta}=\bxi},
\ee
and
\be
\left.\dfrac{\partial \chi_{\rm s}^{\bm \zeta}\left[v_{\rm s}^{\bxi}\right](\br,\br')}{\partial \xi_\lambda}\right|_{{\bm \zeta}=\bxi}=\int d\br''\;\chi_{\rm s}^{[2]\bxi}(\br,\br',\br'')\dfrac{\partial v_{\rm s}^{\bxi}(\br'')}{\partial \xi_\lambda},
\ee
or, in a more compact form,
\be
\left.\dfrac{\partial \chi_{\rm s}^{\bm \zeta}\left[v_{\rm s}^{\bxi}\right]}{\partial \xi_\lambda}\right|_{{\bm \zeta}=\bxi}=\left(\chi_{\rm s}^{[2]\bxi}\star\dfrac{\partial v_{\rm s}^{\bxi}}{\partial \xi_\lambda}\right).
\ee
Note the parentheses on the right-hand side of the above equation that indicate integration over the last position in the ensemble KS quadratic response function, whose sum-over-states expression is derived in Appendix~\ref{app:SOS_KS_quadratic_rsp_func}, for completeness. From Eq.~(\ref{eq:true_ens_KS_pot}) it comes
\be
\dfrac{\partial \chi_{\rm s}^{\bxi}}{\partial \xi_\lambda}=\chi^{\bxi}_{{\rm s},\lambda}-\chi^{\bxi}_{{\rm s},0}+\left(\chi_{\rm s}^{[2]\bxi}\star \dfrac{\partial v^{\bxi}_{\rm Hxc}}{\partial\xi_\lambda}\right).
\ee
Combining the above equation with Eqs.~(\ref{eq:inv_rsp_compt_starting_eq_from_linearity}) and (\ref{eq:deriv_inv_ens_rsp_KS}) leads to 
\begin{equation}
\begin{split}
(R^{\bxi}_\nu)^{-1}&=\left(\chi_{\rm s}^{\bxi}\right)^{-1}+\sum_{\lambda>0}\left(\delta_{\lambda\nu}-\xi_\lambda\right)\left((R^{\bxi}_{{\rm s},\lambda})^{-1}-(R^{\bxi}_{{\rm s},0})^{-1}\right)
\\
&\quad
+\left(\chi_{\rm s}^{\bxi}\right)^{-1}\star\left(\chi_{\rm s}^{[2]\bxi}\star\sum_{\lambda>0}\left(\delta_{\lambda\nu}-\xi_\lambda\right)\dfrac{\partial v^{\bxi}_{\rm Hxc}}{\partial\xi_\lambda}\right)\star \left(\chi_{\rm s}^{\bxi}\right)^{-1}
\\
&\quad
-f_{\rm Hxc}^{\bxi}+
\sum_{\lambda>0}\left(\delta_{\lambda\nu}-\xi_\lambda\right)\dfrac{\partial f_{\rm Hxc}^{\bxi}}{\partial \xi_\lambda}
.
\end{split}
\end{equation}
The Dyson-like Eq.~(\ref{eq:dysonlike}) combined with the individual kernel expression of Eq.~(\ref{eq:Xinu2}) is finally recovered once the following simplification has been made,
\begin{equation}
\begin{split}
&\left(\chi_{\rm s}^{\bxi}\right)^{-1}+\sum_{\lambda>0}\left(\delta_{\lambda\nu}-\xi_\lambda\right)\left((R^{\bxi}_{{\rm s},\lambda})^{-1}-(R^{\bxi}_{{\rm s},0})^{-1}\right)
\\
&=\left(\chi_{\rm s}^{\bxi}\right)^{-1}\left(\chi_{\rm s}^{\bxi}+\sum_{\lambda>0}\left(\delta_{\lambda\nu}-\xi_\lambda\right)\left(\chi^{\bxi}_{{\rm s},\lambda}-\chi^{\bxi}_{{\rm s},0}\right)\right)\left(\chi_{\rm s}^{\bxi}\right)^{-1}
\\
&=\left(\chi_{\rm s}^{\bxi}\right)^{-1}\left(\chi^{\bxi}_{{\rm s},0}+\sum_{\lambda>0}\delta_{\lambda\nu}\left(\chi^{\bxi}_{{\rm s},\lambda}-\chi^{\bxi}_{{\rm s},0}\right)\right)\left(\chi_{\rm s}^{\bxi}\right)^{-1}
\\
&\underset{\nu\geq 0}{=}\left(\chi_{\rm s}^{\bxi}\right)^{-1}\chi^{\bxi}_{{\rm s},\nu}\left(\chi_{\rm s}^{\bxi}\right)^{-1}
\\
&=(R^{\bxi}_{{\rm s},\nu})^{-1}.
\end{split}
\end{equation}

\bibliography{biblio}

\newcommand{\Aa}[0]{Aa}
\providecommand{\latin}[1]{#1}
\makeatletter
\providecommand{\doi}
  {\begingroup\let\do\@makeother\dospecials
  \catcode`\{=1 \catcode`\}=2 \doi@aux}
\providecommand{\doi@aux}[1]{\endgroup\texttt{#1}}
\makeatother
\providecommand*\mcitethebibliography{\thebibliography}
\csname @ifundefined\endcsname{endmcitethebibliography}
  {\let\endmcitethebibliography\endthebibliography}{}
\begin{mcitethebibliography}{101}
\providecommand*\natexlab[1]{#1}
\providecommand*\mciteSetBstSublistMode[1]{}
\providecommand*\mciteSetBstMaxWidthForm[2]{}
\providecommand*\mciteBstWouldAddEndPuncttrue
  {\def\EndOfBibitem{\unskip.}}
\providecommand*\mciteBstWouldAddEndPunctfalse
  {\let\EndOfBibitem\relax}
\providecommand*\mciteSetBstMidEndSepPunct[3]{}
\providecommand*\mciteSetBstSublistLabelBeginEnd[3]{}
\providecommand*\EndOfBibitem{}
\mciteSetBstSublistMode{f}
\mciteSetBstMaxWidthForm{subitem}{(\alph{mcitesubitemcount})}
\mciteSetBstSublistLabelBeginEnd
  {\mcitemaxwidthsubitemform\space}
  {\relax}
  {\relax}

\bibitem[Teale \latin{et~al.}(2022)Teale, Helgaker, Savin, Adamo, Aradi,
  Arbuznikov, Ayers, Baerends, Barone, Calaminici, and {et {\it
  al.}}]{Teale2022_DFT_exchange}
Teale,~A.~M.; Helgaker,~T.; Savin,~A.; Adamo,~C.; Aradi,~B.; Arbuznikov,~A.~V.;
  Ayers,~P.~W.; Baerends,~E.~J.; Barone,~V.; Calaminici,~P. \latin{et~al.}  DFT
  exchange: sharing perspectives on the workhorse of quantum chemistry and
  materials science. \emph{Phys. Chem. Chem. Phys.} \textbf{2022}, \emph{24},
  28700--28781\relax
\mciteBstWouldAddEndPuncttrue
\mciteSetBstMidEndSepPunct{\mcitedefaultmidpunct}
{\mcitedefaultendpunct}{\mcitedefaultseppunct}\relax
\EndOfBibitem
\bibitem[Hohenberg and Kohn(1964)Hohenberg, and Kohn]{hktheo}
Hohenberg,~P.; Kohn,~W. Inhomogeneous Electron Gas. \emph{Phys. Rev.}
  \textbf{1964}, \emph{136}, B864\relax
\mciteBstWouldAddEndPuncttrue
\mciteSetBstMidEndSepPunct{\mcitedefaultmidpunct}
{\mcitedefaultendpunct}{\mcitedefaultseppunct}\relax
\EndOfBibitem
\bibitem[Kohn and Sham(1965)Kohn, and Sham]{KS}
Kohn,~W.; Sham,~L. Self-Consistent Equations Including Exchange and Correlation
  Effects. \emph{Phys. Rev.} \textbf{1965}, \emph{140}, A1133\relax
\mciteBstWouldAddEndPuncttrue
\mciteSetBstMidEndSepPunct{\mcitedefaultmidpunct}
{\mcitedefaultendpunct}{\mcitedefaultseppunct}\relax
\EndOfBibitem
\bibitem[Perdew \latin{et~al.}(1996)Perdew, Burke, and
  Ernzerhof]{perdew1996generalized}
Perdew,~J.~P.; Burke,~K.; Ernzerhof,~M. Generalized gradient approximation made
  simple. \emph{Phys. Rev. Lett.} \textbf{1996}, \emph{77}, 3865\relax
\mciteBstWouldAddEndPuncttrue
\mciteSetBstMidEndSepPunct{\mcitedefaultmidpunct}
{\mcitedefaultendpunct}{\mcitedefaultseppunct}\relax
\EndOfBibitem
\bibitem[Burke(2012)]{Burke2012perspectiveDFA}
Burke,~K. Perspective on density functional theory. \emph{The Journal of
  Chemical Physics} \textbf{2012}, \emph{136}, 150901\relax
\mciteBstWouldAddEndPuncttrue
\mciteSetBstMidEndSepPunct{\mcitedefaultmidpunct}
{\mcitedefaultendpunct}{\mcitedefaultseppunct}\relax
\EndOfBibitem
\bibitem[Goerigk \latin{et~al.}(2017)Goerigk, Hansen, Bauer, Ehrlich, Najibi,
  and Grimme]{Goerigk2017dftzoo}
Goerigk,~L.; Hansen,~A.; Bauer,~C.; Ehrlich,~S.; Najibi,~A.; Grimme,~S. A look
  at the density functional theory zoo with the advanced GMTKN55 database for
  general main group thermochemistry{,} kinetics and noncovalent interactions.
  \emph{Phys. Chem. Chem. Phys.} \textbf{2017}, \emph{19}, 32184--32215\relax
\mciteBstWouldAddEndPuncttrue
\mciteSetBstMidEndSepPunct{\mcitedefaultmidpunct}
{\mcitedefaultendpunct}{\mcitedefaultseppunct}\relax
\EndOfBibitem
\bibitem[Tanaka \latin{et~al.}(2024)Tanaka, Saito, Murata, Nakata, and
  Miyazaki]{Tanaka2024largescaleDFT}
Tanaka,~A.; Saito,~A.; Murata,~T.; Nakata,~A.; Miyazaki,~T. Large-scale DFT
  calculations of multi-component glass systems (SiO2)0.70(Al2O3)0.13(XO)0.17
  (X = Mg, Ca, Sr, Ba) : Accuracy of classical force fields. \emph{Journal of
  Non-Crystalline Solids} \textbf{2024}, \emph{625}, 122714\relax
\mciteBstWouldAddEndPuncttrue
\mciteSetBstMidEndSepPunct{\mcitedefaultmidpunct}
{\mcitedefaultendpunct}{\mcitedefaultseppunct}\relax
\EndOfBibitem
\bibitem[G\"orling(1999)]{Gorling99_Density}
G\"orling,~A. Density-functional theory beyond the Hohenberg-Kohn theorem.
  \emph{Phys. Rev. A} \textbf{1999}, \emph{59}, 3359--3374\relax
\mciteBstWouldAddEndPuncttrue
\mciteSetBstMidEndSepPunct{\mcitedefaultmidpunct}
{\mcitedefaultendpunct}{\mcitedefaultseppunct}\relax
\EndOfBibitem
\bibitem[Yang and Ayers(2024)Yang, and
  Ayers]{yang2024foundationdeltascfapproachdensity}
Yang,~W.; Ayers,~P.~W. Foundation for the {$\Delta$}SCF Approach in Density
  Functional Theory. \emph{arXiv preprint physics.chem-ph} \textbf{2024},
  \url{https://arxiv.org/abs/2403.04604} (accessed 2025-08-18)\relax
\mciteBstWouldAddEndPuncttrue
\mciteSetBstMidEndSepPunct{\mcitedefaultmidpunct}
{\mcitedefaultendpunct}{\mcitedefaultseppunct}\relax
\EndOfBibitem
\bibitem[Runge and Gross(1984)Runge, and Gross]{runge1984density}
Runge,~E.; Gross,~E.~K. Density-functional theory for time-dependent systems.
  \emph{Phys. Rev. Lett.} \textbf{1984}, \emph{52}, 997\relax
\mciteBstWouldAddEndPuncttrue
\mciteSetBstMidEndSepPunct{\mcitedefaultmidpunct}
{\mcitedefaultendpunct}{\mcitedefaultseppunct}\relax
\EndOfBibitem
\bibitem[Casida(1995)]{casida1995timedependent}
Casida,~M.~E. In \emph{Recent Advances in Density Functional Methods};
  Chong,~D.~P., Ed.; World Scientific, 1995; Vol.~1; pp 155--192\relax
\mciteBstWouldAddEndPuncttrue
\mciteSetBstMidEndSepPunct{\mcitedefaultmidpunct}
{\mcitedefaultendpunct}{\mcitedefaultseppunct}\relax
\EndOfBibitem
\bibitem[Casida and Huix-Rotllant(2012)Casida, and
  Huix-Rotllant]{Casida_tddft_review_2012}
Casida,~M.; Huix-Rotllant,~M. Progress in Time-Dependent Density-Functional
  Theory. \emph{Annu. Rev. Phys. Chem.} \textbf{2012}, \emph{63}, 287\relax
\mciteBstWouldAddEndPuncttrue
\mciteSetBstMidEndSepPunct{\mcitedefaultmidpunct}
{\mcitedefaultendpunct}{\mcitedefaultseppunct}\relax
\EndOfBibitem
\bibitem[Lacombe and Maitra(2023)Lacombe, and
  Maitra]{Lacombe2023_Non-adiabatic}
Lacombe,~L.; Maitra,~N. Non-adiabatic approximations in time-dependent density
  functional theory: progress and prospects. \emph{npj Comput Mater}
  \textbf{2023}, \emph{9}, 124\relax
\mciteBstWouldAddEndPuncttrue
\mciteSetBstMidEndSepPunct{\mcitedefaultmidpunct}
{\mcitedefaultendpunct}{\mcitedefaultseppunct}\relax
\EndOfBibitem
\bibitem[Maitra \latin{et~al.}(2004)Maitra, Zhang, Cave, and
  Burke]{maitra2004double}
Maitra,~N.~T.; Zhang,~F.; Cave,~R.~J.; Burke,~K. Double excitations within
  time-dependent density functional theory linear response. \emph{J. Chem.
  Phys.} \textbf{2004}, \emph{120}, 5932\relax
\mciteBstWouldAddEndPuncttrue
\mciteSetBstMidEndSepPunct{\mcitedefaultmidpunct}
{\mcitedefaultendpunct}{\mcitedefaultseppunct}\relax
\EndOfBibitem
\bibitem[Cave \latin{et~al.}(2004)Cave, Zhang, Maitra, and
  Burke]{cave2004dressed}
Cave,~R.~J.; Zhang,~F.; Maitra,~N.~T.; Burke,~K. A dressed TDDFT treatment of
  the 21Ag states of butadiene and hexatriene. \emph{Chem. Phys. Lett.}
  \textbf{2004}, \emph{389}, 39--42\relax
\mciteBstWouldAddEndPuncttrue
\mciteSetBstMidEndSepPunct{\mcitedefaultmidpunct}
{\mcitedefaultendpunct}{\mcitedefaultseppunct}\relax
\EndOfBibitem
\bibitem[Huix-Rotllant \latin{et~al.}(2011)Huix-Rotllant, Ipatov, Rubio, and
  Casida]{Huix-Rotllant2011_Assessment}
Huix-Rotllant,~M.; Ipatov,~A.; Rubio,~A.; Casida,~M.~E. Assessment of dressed
  time-dependent density-functional theory for the low-lying valence states of
  28 organic chromophores. \emph{Chemical Physics} \textbf{2011}, \emph{391},
  120--129, Open problems and new solutions in time dependent density
  functional theory\relax
\mciteBstWouldAddEndPuncttrue
\mciteSetBstMidEndSepPunct{\mcitedefaultmidpunct}
{\mcitedefaultendpunct}{\mcitedefaultseppunct}\relax
\EndOfBibitem
\bibitem[Elliott \latin{et~al.}(2011)Elliott, Goldson, Canahui, and
  Maitra]{elliott2011perspectives}
Elliott,~P.; Goldson,~S.; Canahui,~C.; Maitra,~N.~T. Perspectives on
  double-excitations in TDDFT. \emph{Chem. Phys.} \textbf{2011}, \emph{391},
  110--119, Open problems and new solutions in time dependent density
  functional theory\relax
\mciteBstWouldAddEndPuncttrue
\mciteSetBstMidEndSepPunct{\mcitedefaultmidpunct}
{\mcitedefaultendpunct}{\mcitedefaultseppunct}\relax
\EndOfBibitem
\bibitem[Maitra(2022)]{maitra2022double}
Maitra,~N.~T. Double and Charge-Transfer Excitations in Time-Dependent Density
  Functional Theory. \emph{Annu. Rev. Phys. Chem.} \textbf{2022}, \emph{73},
  117--140\relax
\mciteBstWouldAddEndPuncttrue
\mciteSetBstMidEndSepPunct{\mcitedefaultmidpunct}
{\mcitedefaultendpunct}{\mcitedefaultseppunct}\relax
\EndOfBibitem
\bibitem[Matsika(2021)]{Matsika2021elstructCI}
Matsika,~S. Electronic Structure Methods for the Description of Nonadiabatic
  Effects and Conical Intersections. \emph{Chemical Reviews} \textbf{2021},
  \emph{121}, 9407--9449, PMID: 34156838\relax
\mciteBstWouldAddEndPuncttrue
\mciteSetBstMidEndSepPunct{\mcitedefaultmidpunct}
{\mcitedefaultendpunct}{\mcitedefaultseppunct}\relax
\EndOfBibitem
\bibitem[Dar and Maitra(2023)Dar, and Maitra]{Dar2023_TDDFT_DE}
Dar,~D.~B.; Maitra,~N.~T. Oscillator strengths and excited-state couplings for
  double excitations in time-dependent density functional theory. \emph{The
  Journal of Chemical Physics} \textbf{2023}, \emph{159}, 211104\relax
\mciteBstWouldAddEndPuncttrue
\mciteSetBstMidEndSepPunct{\mcitedefaultmidpunct}
{\mcitedefaultendpunct}{\mcitedefaultseppunct}\relax
\EndOfBibitem
\bibitem[Dar and Maitra(2025)Dar, and Maitra]{Dar2025_DTDDFT_CI}
Dar,~D.~B.; Maitra,~N.~T. Capturing the Elusive Curve-Crossing in Low-Lying
  States of Butadiene with Dressed TDDFT. \emph{The Journal of Physical
  Chemistry Letters} \textbf{2025}, \emph{16}, 703--709, PMID: 39792779\relax
\mciteBstWouldAddEndPuncttrue
\mciteSetBstMidEndSepPunct{\mcitedefaultmidpunct}
{\mcitedefaultendpunct}{\mcitedefaultseppunct}\relax
\EndOfBibitem
\bibitem[Baranova and Maitra(2025)Baranova, and
  Maitra]{baranova2025excitedstatedensitiestimedependent}
Baranova,~A.; Maitra,~N.~T. Excited State Densities from Time-Dependent Density
  Functional Response Theory. \emph{arXiv preprint physics.chem-ph}
  \textbf{2025}, \url{https://arxiv.org/abs/2506.05082} (accessed
  2025-08-18)\relax
\mciteBstWouldAddEndPuncttrue
\mciteSetBstMidEndSepPunct{\mcitedefaultmidpunct}
{\mcitedefaultendpunct}{\mcitedefaultseppunct}\relax
\EndOfBibitem
\bibitem[Fan(1949)]{Fan1949_On}
Fan,~K. On a theorem of Weyl concerning eigenvalues of linear transformations
  I. \emph{PNAS} \textbf{1949}, \emph{35}, 652\relax
\mciteBstWouldAddEndPuncttrue
\mciteSetBstMidEndSepPunct{\mcitedefaultmidpunct}
{\mcitedefaultendpunct}{\mcitedefaultseppunct}\relax
\EndOfBibitem
\bibitem[Theophilou(1979)]{JPC79_Theophilou_equi-ensembles}
Theophilou,~A.~K. The energy density functional formalism for excited states.
  \emph{J. Phys. C: Solid State Phys.} \textbf{1979}, \emph{12}, 5419\relax
\mciteBstWouldAddEndPuncttrue
\mciteSetBstMidEndSepPunct{\mcitedefaultmidpunct}
{\mcitedefaultendpunct}{\mcitedefaultseppunct}\relax
\EndOfBibitem
\bibitem[Hendekovic(1982)]{Hendekovic1982_equi-ensembles}
Hendekovic,~J. On the energy variation method. \emph{Chemical Physics Letters}
  \textbf{1982}, \emph{90}, 198--201\relax
\mciteBstWouldAddEndPuncttrue
\mciteSetBstMidEndSepPunct{\mcitedefaultmidpunct}
{\mcitedefaultendpunct}{\mcitedefaultseppunct}\relax
\EndOfBibitem
\bibitem[Gross \latin{et~al.}(1988)Gross, Oliveira, and
  Kohn]{gross1988rayleigh}
Gross,~E. K.~U.; Oliveira,~L.~N.; Kohn,~W. Rayleigh-Ritz variational principle
  for ensembles of fractionally occupied states. \emph{Phys. Rev. A}
  \textbf{1988}, \emph{37}, 2805\relax
\mciteBstWouldAddEndPuncttrue
\mciteSetBstMidEndSepPunct{\mcitedefaultmidpunct}
{\mcitedefaultendpunct}{\mcitedefaultseppunct}\relax
\EndOfBibitem
\bibitem[Gross \latin{et~al.}(1988)Gross, Oliveira, and Kohn]{gross1988density}
Gross,~E. K.~U.; Oliveira,~L.~N.; Kohn,~W. Density-functional theory for
  ensembles of fractionally occupied states. I. Basic formalism. \emph{Phys.
  Rev. A} \textbf{1988}, \emph{37}, 2809\relax
\mciteBstWouldAddEndPuncttrue
\mciteSetBstMidEndSepPunct{\mcitedefaultmidpunct}
{\mcitedefaultendpunct}{\mcitedefaultseppunct}\relax
\EndOfBibitem
\bibitem[Oliveira \latin{et~al.}(1988)Oliveira, Gross, and
  Kohn]{oliveira1988density}
Oliveira,~L.~N.; Gross,~E. K.~U.; Kohn,~W. Density-functional theory for
  ensembles of fractionally occupied states. II. Application to the He atom.
  \emph{Phys. Rev. A} \textbf{1988}, \emph{37}, 2821\relax
\mciteBstWouldAddEndPuncttrue
\mciteSetBstMidEndSepPunct{\mcitedefaultmidpunct}
{\mcitedefaultendpunct}{\mcitedefaultseppunct}\relax
\EndOfBibitem
\bibitem[Deur \latin{et~al.}(2017)Deur, Mazouin, and Fromager]{deur2017exact}
Deur,~K.; Mazouin,~L.; Fromager,~E. Exact ensemble density functional theory
  for excited states in a model system: Investigating the weight dependence of
  the correlation energy. \emph{Phys. Rev. B} \textbf{2017}, \emph{95},
  035120\relax
\mciteBstWouldAddEndPuncttrue
\mciteSetBstMidEndSepPunct{\mcitedefaultmidpunct}
{\mcitedefaultendpunct}{\mcitedefaultseppunct}\relax
\EndOfBibitem
\bibitem[Yang \latin{et~al.}(2017)Yang, Pribram-Jones, Burke, and
  Ullrich]{yang2017direct}
Yang,~Z.-h.; Pribram-Jones,~A.; Burke,~K.; Ullrich,~C.~A. Direct extraction of
  excitation energies from ensemble density-functional theory. \emph{Phys. Rev.
  Lett.} \textbf{2017}, \emph{119}, 033003\relax
\mciteBstWouldAddEndPuncttrue
\mciteSetBstMidEndSepPunct{\mcitedefaultmidpunct}
{\mcitedefaultendpunct}{\mcitedefaultseppunct}\relax
\EndOfBibitem
\bibitem[Gould and Pittalis(2017)Gould, and Pittalis]{gould2017hartree}
Gould,~T.; Pittalis,~S. Hartree and Exchange in Ensemble Density Functional
  Theory: Avoiding the Nonuniqueness Disaster. \emph{Phys. Rev. Lett.}
  \textbf{2017}, \emph{119}, 243001\relax
\mciteBstWouldAddEndPuncttrue
\mciteSetBstMidEndSepPunct{\mcitedefaultmidpunct}
{\mcitedefaultendpunct}{\mcitedefaultseppunct}\relax
\EndOfBibitem
\bibitem[Gould \latin{et~al.}(2018)Gould, Kronik, and
  Pittalis]{gould2018charge}
Gould,~T.; Kronik,~L.; Pittalis,~S. Charge transfer excitations from exact and
  approximate ensemble Kohn-Sham theory. \emph{J. Chem. Phys.} \textbf{2018},
  \emph{148}, 174101\relax
\mciteBstWouldAddEndPuncttrue
\mciteSetBstMidEndSepPunct{\mcitedefaultmidpunct}
{\mcitedefaultendpunct}{\mcitedefaultseppunct}\relax
\EndOfBibitem
\bibitem[Deur \latin{et~al.}(2018)Deur, Mazouin, Senjean, and
  Fromager]{deur2018exploring}
Deur,~K.; Mazouin,~L.; Senjean,~B.; Fromager,~E. Exploring weight-dependent
  density-functional approximations for ensembles in the Hubbard dimer.
  \emph{Eur. Phys. J. B} \textbf{2018}, \emph{91}, 162\relax
\mciteBstWouldAddEndPuncttrue
\mciteSetBstMidEndSepPunct{\mcitedefaultmidpunct}
{\mcitedefaultendpunct}{\mcitedefaultseppunct}\relax
\EndOfBibitem
\bibitem[Gould and Pittalis(2019)Gould, and
  Pittalis]{PRL19_Gould_DD_correlation}
Gould,~T.; Pittalis,~S. Density-Driven Correlations in Many-Electron Ensembles:
  Theory and Application for Excited States. \emph{Phys. Rev. Lett.}
  \textbf{2019}, \emph{123}, 016401\relax
\mciteBstWouldAddEndPuncttrue
\mciteSetBstMidEndSepPunct{\mcitedefaultmidpunct}
{\mcitedefaultendpunct}{\mcitedefaultseppunct}\relax
\EndOfBibitem
\bibitem[Fromager(2020)]{Fromager_2020}
Fromager,~E. Individual Correlations in Ensemble Density Functional Theory:
  State- and Density-Driven Decompositions without Additional Kohn-Sham
  Systems. \emph{Phys. Rev. Lett.} \textbf{2020}, \emph{124}, 243001\relax
\mciteBstWouldAddEndPuncttrue
\mciteSetBstMidEndSepPunct{\mcitedefaultmidpunct}
{\mcitedefaultendpunct}{\mcitedefaultseppunct}\relax
\EndOfBibitem
\bibitem[Gould \latin{et~al.}(2020)Gould, Stefanucci, and
  Pittalis]{PRL20_Gould_Hartree_def_from_ACDF_th}
Gould,~T.; Stefanucci,~G.; Pittalis,~S. Ensemble Density Functional Theory:
  Insight from the Fluctuation-Dissipation Theorem. \emph{Phys. Rev. Lett.}
  \textbf{2020}, \emph{125}, 233001\relax
\mciteBstWouldAddEndPuncttrue
\mciteSetBstMidEndSepPunct{\mcitedefaultmidpunct}
{\mcitedefaultendpunct}{\mcitedefaultseppunct}\relax
\EndOfBibitem
\bibitem[Gould(2020)]{Gould2020_Approximately}
Gould,~T. Approximately Self-Consistent Ensemble Density Functional Theory:
  Toward Inclusion of All Correlations. \emph{J. Phys. Chem. Lett.}
  \textbf{2020}, \emph{11}, 9907--9912\relax
\mciteBstWouldAddEndPuncttrue
\mciteSetBstMidEndSepPunct{\mcitedefaultmidpunct}
{\mcitedefaultendpunct}{\mcitedefaultseppunct}\relax
\EndOfBibitem
\bibitem[Loos and Fromager(2020)Loos, and Fromager]{loos2020weightdependent}
Loos,~P.-F.; Fromager,~E. A weight-dependent local correlation
  density-functional approximation for ensembles. \emph{J. Chem. Phys.}
  \textbf{2020}, \emph{152}, 214101\relax
\mciteBstWouldAddEndPuncttrue
\mciteSetBstMidEndSepPunct{\mcitedefaultmidpunct}
{\mcitedefaultendpunct}{\mcitedefaultseppunct}\relax
\EndOfBibitem
\bibitem[Yang(2021)]{Yang2021_Second}
Yang,~Z.-h. Second-order perturbative correlation energy functional in the
  ensemble density-functional theory. \emph{Phys. Rev. A} \textbf{2021},
  \emph{104}, 052806\relax
\mciteBstWouldAddEndPuncttrue
\mciteSetBstMidEndSepPunct{\mcitedefaultmidpunct}
{\mcitedefaultendpunct}{\mcitedefaultseppunct}\relax
\EndOfBibitem
\bibitem[Gould and Kronik(2021)Gould, and Kronik]{Gould2021_Ensemble_ugly}
Gould,~T.; Kronik,~L. {Ensemble generalized Kohn--Sham theory: The good, the
  bad, and the ugly}. \emph{J. Chem. Phys.} \textbf{2021}, \emph{154},
  094125\relax
\mciteBstWouldAddEndPuncttrue
\mciteSetBstMidEndSepPunct{\mcitedefaultmidpunct}
{\mcitedefaultendpunct}{\mcitedefaultseppunct}\relax
\EndOfBibitem
\bibitem[Gould and Pittalis(2024)Gould, and Pittalis]{gould2024local}
Gould,~T.; Pittalis,~S. Local Density Approximation for Excited States.
  \emph{Phys. Rev. X} \textbf{2024}, \emph{14}, 041045\relax
\mciteBstWouldAddEndPuncttrue
\mciteSetBstMidEndSepPunct{\mcitedefaultmidpunct}
{\mcitedefaultendpunct}{\mcitedefaultseppunct}\relax
\EndOfBibitem
\bibitem[Gould \latin{et~al.}(2023)Gould, Kooi, Gori-Giorgi, and
  Pittalis]{Gould2023_Electronic}
Gould,~T.; Kooi,~D.~P.; Gori-Giorgi,~P.; Pittalis,~S. Electronic Excited States
  in Extreme Limits via Ensemble Density Functionals. \emph{Phys. Rev. Lett.}
  \textbf{2023}, \emph{130}, 106401\relax
\mciteBstWouldAddEndPuncttrue
\mciteSetBstMidEndSepPunct{\mcitedefaultmidpunct}
{\mcitedefaultendpunct}{\mcitedefaultseppunct}\relax
\EndOfBibitem
\bibitem[Gould \latin{et~al.}(2022)Gould, Hashimi, Kronik, and
  Dale]{gould2022single}
Gould,~T.; Hashimi,~Z.; Kronik,~L.; Dale,~S.~G. Single Excitation Energies
  Obtained from the Ensemble "HOMO-LUMO Gap": Exact Results and Approximations.
  \emph{J. Phys. Chem. Lett.} \textbf{2022}, \emph{13}, 2452--2458\relax
\mciteBstWouldAddEndPuncttrue
\mciteSetBstMidEndSepPunct{\mcitedefaultmidpunct}
{\mcitedefaultendpunct}{\mcitedefaultseppunct}\relax
\EndOfBibitem
\bibitem[Gould \latin{et~al.}(2021)Gould, Kronik, and
  Pittalis]{Gould2021_Double}
Gould,~T.; Kronik,~L.; Pittalis,~S. Double excitations in molecules from
  ensemble density functionals: Theory and approximations. \emph{Phys. Rev. A}
  \textbf{2021}, \emph{104}, 022803\relax
\mciteBstWouldAddEndPuncttrue
\mciteSetBstMidEndSepPunct{\mcitedefaultmidpunct}
{\mcitedefaultendpunct}{\mcitedefaultseppunct}\relax
\EndOfBibitem
\bibitem[Cernatic \latin{et~al.}(2022)Cernatic, Senjean, Robert, and
  Fromager]{Cernatic2022}
Cernatic,~F.; Senjean,~B.; Robert,~V.; Fromager,~E. Ensemble Density Functional
  Theory of Neutral and Charged Excitations. \emph{Top Curr Chem (Z)}
  \textbf{2022}, \emph{380}, 4\relax
\mciteBstWouldAddEndPuncttrue
\mciteSetBstMidEndSepPunct{\mcitedefaultmidpunct}
{\mcitedefaultendpunct}{\mcitedefaultseppunct}\relax
\EndOfBibitem
\bibitem[Schilling and Pittalis(2021)Schilling, and
  Pittalis]{Schilling2021_Ensemble}
Schilling,~C.; Pittalis,~S. Ensemble Reduced Density Matrix Functional Theory
  for Excited States and Hierarchical Generalization of Pauli's Exclusion
  Principle. \emph{Phys. Rev. Lett.} \textbf{2021}, \emph{127}, 023001\relax
\mciteBstWouldAddEndPuncttrue
\mciteSetBstMidEndSepPunct{\mcitedefaultmidpunct}
{\mcitedefaultendpunct}{\mcitedefaultseppunct}\relax
\EndOfBibitem
\bibitem[Liebert \latin{et~al.}(2022)Liebert, Castillo, Labb{\'e}, and
  Schilling]{Liebert2022_Foundation}
Liebert,~J.; Castillo,~F.; Labb{\'e},~J.-P.; Schilling,~C. Foundation of
  One-Particle Reduced Density Matrix Functional Theory for Excited States.
  \emph{J. Chem. Theory Comput.} \textbf{2022}, \emph{18}, 124--140\relax
\mciteBstWouldAddEndPuncttrue
\mciteSetBstMidEndSepPunct{\mcitedefaultmidpunct}
{\mcitedefaultendpunct}{\mcitedefaultseppunct}\relax
\EndOfBibitem
\bibitem[Benavides-Riveros \latin{et~al.}(2022)Benavides-Riveros, Chen,
  Schilling, Mantilla, and Pittalis]{Benavides-Riveros2022_Excitations}
Benavides-Riveros,~C.~L.; Chen,~L.; Schilling,~C.; Mantilla,~S.; Pittalis,~S.
  Excitations of Quantum Many-Body Systems via Purified Ensembles: A
  Unitary-Coupled-Cluster-Based Approach. \emph{Phys. Rev. Lett.}
  \textbf{2022}, \emph{129}, 066401\relax
\mciteBstWouldAddEndPuncttrue
\mciteSetBstMidEndSepPunct{\mcitedefaultmidpunct}
{\mcitedefaultendpunct}{\mcitedefaultseppunct}\relax
\EndOfBibitem
\bibitem[Liebert and Schilling(2023)Liebert, and
  Schilling]{Liebert_2023_An_exact_bosons}
Liebert,~J.; Schilling,~C. An exact one-particle theory of bosonic excitations:
  from a generalized Hohenberg--Kohn theorem to convexified N-representability.
  \emph{New J. Phys.} \textbf{2023}, \emph{25}, 013009\relax
\mciteBstWouldAddEndPuncttrue
\mciteSetBstMidEndSepPunct{\mcitedefaultmidpunct}
{\mcitedefaultendpunct}{\mcitedefaultseppunct}\relax
\EndOfBibitem
\bibitem[Liebert and Schilling(2023)Liebert, and
  Schilling]{Liebert2023_Deriving}
Liebert,~J.; Schilling,~C. {Deriving density-matrix functionals for excited
  states}. \emph{SciPost Phys.} \textbf{2023}, \emph{14}, 120\relax
\mciteBstWouldAddEndPuncttrue
\mciteSetBstMidEndSepPunct{\mcitedefaultmidpunct}
{\mcitedefaultendpunct}{\mcitedefaultseppunct}\relax
\EndOfBibitem
\bibitem[Ding \latin{et~al.}(2024)Ding, Hong, and Schilling]{ding2024ground}
Ding,~L.; Hong,~C.-L.; Schilling,~C. Ground and Excited States from Ensemble
  Variational Principles. \emph{arXiv preprint quant-ph} \textbf{2024},
  \url{https://arxiv.org/abs/2401.12104} (accessed 2025-08-18)\relax
\mciteBstWouldAddEndPuncttrue
\mciteSetBstMidEndSepPunct{\mcitedefaultmidpunct}
{\mcitedefaultendpunct}{\mcitedefaultseppunct}\relax
\EndOfBibitem
\bibitem[Scott \latin{et~al.}(2024)Scott, Kozlowski, Crisostomo, Pribram-Jones,
  and Burke]{Scott2024_Exact}
Scott,~T.~R.; Kozlowski,~J.; Crisostomo,~S.; Pribram-Jones,~A.; Burke,~K. Exact
  conditions for ensemble density functional theory. \emph{Phys. Rev. B}
  \textbf{2024}, \emph{109}, 195120\relax
\mciteBstWouldAddEndPuncttrue
\mciteSetBstMidEndSepPunct{\mcitedefaultmidpunct}
{\mcitedefaultendpunct}{\mcitedefaultseppunct}\relax
\EndOfBibitem
\bibitem[Cernatic \latin{et~al.}(2024)Cernatic, Loos, Senjean, and
  Fromager]{Cernatic2024_Neutral}
Cernatic,~F.; Loos,~P.-F.; Senjean,~B.; Fromager,~E. Neutral electronic
  excitations and derivative discontinuities: An extended $N$-centered ensemble
  density functional theory perspective. \emph{Phys. Rev. B} \textbf{2024},
  \emph{109}, 235113\relax
\mciteBstWouldAddEndPuncttrue
\mciteSetBstMidEndSepPunct{\mcitedefaultmidpunct}
{\mcitedefaultendpunct}{\mcitedefaultseppunct}\relax
\EndOfBibitem
\bibitem[Cernatic and Fromager(2024)Cernatic, and
  Fromager]{cernatic2024extended_doubles}
Cernatic,~F.; Fromager,~E. Extended N-centered ensemble density functional
  theory of double electronic excitations. \emph{Journal of Computational
  Chemistry} \textbf{2024}, \emph{45}, 1945--1962\relax
\mciteBstWouldAddEndPuncttrue
\mciteSetBstMidEndSepPunct{\mcitedefaultmidpunct}
{\mcitedefaultendpunct}{\mcitedefaultseppunct}\relax
\EndOfBibitem
\bibitem[Gould \latin{et~al.}(2025)Gould, Dale, Kronik, and
  Pittalis]{Gould2025_PRL_tate-Specific}
Gould,~T.; Dale,~S.~G.; Kronik,~L.; Pittalis,~S. State-Specific Density
  Functionals for Excited States via a Density-Driven Correlation Model.
  \emph{Phys. Rev. Lett.} \textbf{2025}, \emph{134}, 228001\relax
\mciteBstWouldAddEndPuncttrue
\mciteSetBstMidEndSepPunct{\mcitedefaultmidpunct}
{\mcitedefaultendpunct}{\mcitedefaultseppunct}\relax
\EndOfBibitem
\bibitem[Ayers and Levy(2009)Ayers, and Levy]{ayers2009pra}
Ayers,~P.~W.; Levy,~M. Time-independent (static) density-functional theories
  for pure excited states: Extensions and unification. \emph{Phys. Rev. A}
  \textbf{2009}, \emph{80}, 012508\relax
\mciteBstWouldAddEndPuncttrue
\mciteSetBstMidEndSepPunct{\mcitedefaultmidpunct}
{\mcitedefaultendpunct}{\mcitedefaultseppunct}\relax
\EndOfBibitem
\bibitem[Gilbert \latin{et~al.}(2008)Gilbert, Besley, and
  Gill]{Gilbert08_Self-Consistent_MOM}
Gilbert,~A. T.~B.; Besley,~N.~A.; Gill,~P. M.~W. Self-Consistent Field
  Calculations of Excited States Using the Maximum Overlap Method (MOM).
  \emph{The Journal of Physical Chemistry A} \textbf{2008}, \emph{112},
  13164--13171\relax
\mciteBstWouldAddEndPuncttrue
\mciteSetBstMidEndSepPunct{\mcitedefaultmidpunct}
{\mcitedefaultendpunct}{\mcitedefaultseppunct}\relax
\EndOfBibitem
\bibitem[Besley \latin{et~al.}(2009)Besley, Gilbert, and
  Gill]{Besley09_Self-consistent_MUM}
Besley,~N.~A.; Gilbert,~A. T.~B.; Gill,~P. M.~W. {Self-consistent-field
  calculations of core excited states}. \emph{The Journal of Chemical Physics}
  \textbf{2009}, \emph{130}, 124308\relax
\mciteBstWouldAddEndPuncttrue
\mciteSetBstMidEndSepPunct{\mcitedefaultmidpunct}
{\mcitedefaultendpunct}{\mcitedefaultseppunct}\relax
\EndOfBibitem
\bibitem[Barca \latin{et~al.}(2018)Barca, Gilbert, and
  Gill]{Barca18_Simple_MOM}
Barca,~G. M.~J.; Gilbert,~A. T.~B.; Gill,~P. M.~W. Simple Models for Difficult
  Electronic Excitations. \emph{Journal of Chemical Theory and Computation}
  \textbf{2018}, \emph{14}, 1501--1509\relax
\mciteBstWouldAddEndPuncttrue
\mciteSetBstMidEndSepPunct{\mcitedefaultmidpunct}
{\mcitedefaultendpunct}{\mcitedefaultseppunct}\relax
\EndOfBibitem
\bibitem[Ziegler \latin{et~al.}(2009)Ziegler, Seth, Krykunov, Autschbach, and
  Wang]{JCP09_Ziegler_relation_TD-DFT_VDFT}
Ziegler,~T.; Seth,~M.; Krykunov,~M.; Autschbach,~J.; Wang,~F. On the relation
  between time-dependent and variational density functional theory approaches
  for the determination of excitation energies and transition moments. \emph{J.
  Chem. Phys.} \textbf{2009}, \emph{130}, 154102\relax
\mciteBstWouldAddEndPuncttrue
\mciteSetBstMidEndSepPunct{\mcitedefaultmidpunct}
{\mcitedefaultendpunct}{\mcitedefaultseppunct}\relax
\EndOfBibitem
\bibitem[Krykunov and Ziegler(2013)Krykunov, and
  Ziegler]{JCTC13_Ziegler_SCF-CV-DFT}
Krykunov,~M.; Ziegler,~T. Self-consistent Formulation of Constricted
  Variational Density Functional Theory with Orbital Relaxation. Implementation
  and Applications. \emph{J. Chem. Theory Comput.} \textbf{2013}, \emph{9},
  2761\relax
\mciteBstWouldAddEndPuncttrue
\mciteSetBstMidEndSepPunct{\mcitedefaultmidpunct}
{\mcitedefaultendpunct}{\mcitedefaultseppunct}\relax
\EndOfBibitem
\bibitem[Glushkov and Levy(2016)Glushkov, and Levy]{levy2016computation}
Glushkov,~V.; Levy,~M. Highly Excited States from a Time Independent Density
  Functional Method. \emph{Computation} \textbf{2016}, \emph{4}, 28\relax
\mciteBstWouldAddEndPuncttrue
\mciteSetBstMidEndSepPunct{\mcitedefaultmidpunct}
{\mcitedefaultendpunct}{\mcitedefaultseppunct}\relax
\EndOfBibitem
\bibitem[Ayers \latin{et~al.}(2012)Ayers, Levy, and
  Nagy]{PRA12_Nagy_TinD-DFT_ES}
Ayers,~P.~W.; Levy,~M.; Nagy,~A. Time-independent density-functional theory for
  excited states of Coulomb systems. \emph{Phys. Rev. A} \textbf{2012},
  \emph{85}, 042518\relax
\mciteBstWouldAddEndPuncttrue
\mciteSetBstMidEndSepPunct{\mcitedefaultmidpunct}
{\mcitedefaultendpunct}{\mcitedefaultseppunct}\relax
\EndOfBibitem
\bibitem[Ayers \latin{et~al.}(2015)Ayers, Levy, and
  Nagy]{JCP15_Ayers_KS-DFT_excit-states_Coulomb}
Ayers,~P.~W.; Levy,~M.; Nagy,~A. Communication: Kohn-Sham theory for excited
  states of Coulomb systems. \emph{J. Chem. Phys.} \textbf{2015}, \emph{143},
  191101\relax
\mciteBstWouldAddEndPuncttrue
\mciteSetBstMidEndSepPunct{\mcitedefaultmidpunct}
{\mcitedefaultendpunct}{\mcitedefaultseppunct}\relax
\EndOfBibitem
\bibitem[Ayers \latin{et~al.}(2018)Ayers, Levy, and
  Nagy]{Ayers2018_Article_Time-independentDensityFunctio}
Ayers,~P.~W.; Levy,~M.; Nagy,~{\'A}. Time-independent density functional theory
  for degenerate excited states of Coulomb systems. \emph{Theor. Chem. Acc.}
  \textbf{2018}, \emph{137}, 152\relax
\mciteBstWouldAddEndPuncttrue
\mciteSetBstMidEndSepPunct{\mcitedefaultmidpunct}
{\mcitedefaultendpunct}{\mcitedefaultseppunct}\relax
\EndOfBibitem
\bibitem[Garrigue(2022)]{Garrigue2022}
Garrigue,~L. Building Kohn--Sham Potentials for Ground and Excited States.
  \emph{Archive for Rational Mechanics and Analysis} \textbf{2022}, \emph{245},
  949--1003\relax
\mciteBstWouldAddEndPuncttrue
\mciteSetBstMidEndSepPunct{\mcitedefaultmidpunct}
{\mcitedefaultendpunct}{\mcitedefaultseppunct}\relax
\EndOfBibitem
\bibitem[Giarrusso and Loos(2023)Giarrusso, and Loos]{Giarrusso2023_Exact}
Giarrusso,~S.; Loos,~P.-F. Exact Excited-State Functionals of the Asymmetric
  Hubbard Dimer. \emph{J. Phys. Chem. Lett.} \textbf{2023}, \emph{14},
  8780--8786\relax
\mciteBstWouldAddEndPuncttrue
\mciteSetBstMidEndSepPunct{\mcitedefaultmidpunct}
{\mcitedefaultendpunct}{\mcitedefaultseppunct}\relax
\EndOfBibitem
\bibitem[Loos and Giarrusso(2025)Loos, and Giarrusso]{Loos2025excistspHD}
Loos,~P.-F.; Giarrusso,~S. Excited-state-specific Kohn–Sham formalism for the
  asymmetric Hubbard dimer. \emph{The Journal of Chemical Physics}
  \textbf{2025}, \emph{162}, 144104\relax
\mciteBstWouldAddEndPuncttrue
\mciteSetBstMidEndSepPunct{\mcitedefaultmidpunct}
{\mcitedefaultendpunct}{\mcitedefaultseppunct}\relax
\EndOfBibitem
\bibitem[Loos(2025)]{Loos2025excstUEG}
Loos,~P.-F. Excited states of the uniform electron gas. \emph{The Journal of
  Chemical Physics} \textbf{2025}, \emph{162}, 204105\relax
\mciteBstWouldAddEndPuncttrue
\mciteSetBstMidEndSepPunct{\mcitedefaultmidpunct}
{\mcitedefaultendpunct}{\mcitedefaultseppunct}\relax
\EndOfBibitem
\bibitem[Levi \latin{et~al.}(2020)Levi, Ivanov, and
  J\'{o}nsson]{Levi20_Variational}
Levi,~G.; Ivanov,~A.~V.; J\'{o}nsson,~H. Variational calculations of excited
  states via direct optimization of the orbitals in DFT. \emph{Faraday
  Discuss.} \textbf{2020}, \emph{224}, 448--466\relax
\mciteBstWouldAddEndPuncttrue
\mciteSetBstMidEndSepPunct{\mcitedefaultmidpunct}
{\mcitedefaultendpunct}{\mcitedefaultseppunct}\relax
\EndOfBibitem
\bibitem[Ivanov \latin{et~al.}(2021)Ivanov, Levi, J\'{o}nsson, and
  J\'{o}nsson]{Ivanov21_Method}
Ivanov,~A.~V.; Levi,~G.; J\'{o}nsson,~E.~O.; J\'{o}nsson,~H. Method for
  Calculating Excited Electronic States Using Density Functionals and Direct
  Orbital Optimization with Real Space Grid or Plane-Wave Basis Set.
  \emph{Journal of Chemical Theory and Computation} \textbf{2021}, \emph{17},
  5034--5049\relax
\mciteBstWouldAddEndPuncttrue
\mciteSetBstMidEndSepPunct{\mcitedefaultmidpunct}
{\mcitedefaultendpunct}{\mcitedefaultseppunct}\relax
\EndOfBibitem
\bibitem[Hait and Head-Gordon(2021)Hait, and Head-Gordon]{Hait21_Orbital}
Hait,~D.; Head-Gordon,~M. Orbital Optimized Density Functional Theory for
  Electronic Excited States. \emph{The Journal of Physical Chemistry Letters}
  \textbf{2021}, \emph{12}, 4517--4529\relax
\mciteBstWouldAddEndPuncttrue
\mciteSetBstMidEndSepPunct{\mcitedefaultmidpunct}
{\mcitedefaultendpunct}{\mcitedefaultseppunct}\relax
\EndOfBibitem
\bibitem[Schmerwitz \latin{et~al.}(2022)Schmerwitz, Ivanov, J\'{o}nsson,
  J\'{o}nsson, and Levi]{Schmerwitz22_Variational}
Schmerwitz,~Y. L.~A.; Ivanov,~A.~V.; J\'{o}nsson,~E.~O.; J\'{o}nsson,~H.;
  Levi,~G. Variational Density Functional Calculations of Excited States:
  Conical Intersection and Avoided Crossing in Ethylene Bond Twisting.
  \emph{The Journal of Physical Chemistry Letters} \textbf{2022}, \emph{13},
  3990--3999\relax
\mciteBstWouldAddEndPuncttrue
\mciteSetBstMidEndSepPunct{\mcitedefaultmidpunct}
{\mcitedefaultendpunct}{\mcitedefaultseppunct}\relax
\EndOfBibitem
\bibitem[Sigurdarson \latin{et~al.}(2023)Sigurdarson, Schmerwitz, Tveiten,
  Levi, and Jónsson]{OODFTRydberg_Levi2023}
Sigurdarson,~A.~E.; Schmerwitz,~Y. L.~A.; Tveiten,~D. K.~V.; Levi,~G.;
  Jónsson,~H. Orbital-optimized density functional calculations of molecular
  Rydberg excited states with real space grid representation and
  self-interaction correction. \emph{The Journal of Chemical Physics}
  \textbf{2023}, \emph{159}, 214109\relax
\mciteBstWouldAddEndPuncttrue
\mciteSetBstMidEndSepPunct{\mcitedefaultmidpunct}
{\mcitedefaultendpunct}{\mcitedefaultseppunct}\relax
\EndOfBibitem
\bibitem[Ivanov \latin{et~al.}(2023)Ivanov, Schmerwitz, Levi, and
  Jónsson]{LeviOODFT_diamond_2023}
Ivanov,~A.~V.; Schmerwitz,~Y. L.~A.; Levi,~G.; Jónsson,~H. {Electronic
  excitations of the charged nitrogen-vacancy center in diamond obtained using
  time-independent variational density functional calculations}. \emph{SciPost
  Phys.} \textbf{2023}, \emph{15}, 009\relax
\mciteBstWouldAddEndPuncttrue
\mciteSetBstMidEndSepPunct{\mcitedefaultmidpunct}
{\mcitedefaultendpunct}{\mcitedefaultseppunct}\relax
\EndOfBibitem
\bibitem[Selenius \latin{et~al.}(2024)Selenius, Sigurdarson, Schmerwitz, and
  Levi]{Levi_OODFTvsTDDFT2024}
Selenius,~E.; Sigurdarson,~A.~E.; Schmerwitz,~Y. L.~A.; Levi,~G.
  Orbital-Optimized Versus Time-Dependent Density Functional Calculations of
  Intramolecular Charge Transfer Excited States. \emph{Journal of Chemical
  Theory and Computation} \textbf{2024}, \emph{20}, 3809--3822, PMID:
  38695313\relax
\mciteBstWouldAddEndPuncttrue
\mciteSetBstMidEndSepPunct{\mcitedefaultmidpunct}
{\mcitedefaultendpunct}{\mcitedefaultseppunct}\relax
\EndOfBibitem
\bibitem[Gould(2025)]{gould2025stationaryconditionsexcitedstates}
Gould,~T. Variational principles in ensemble and excited-state density- and
  potential-functional theories. \emph{Phys. Rev. A} \textbf{2025}, \emph{111},
  032806\relax
\mciteBstWouldAddEndPuncttrue
\mciteSetBstMidEndSepPunct{\mcitedefaultmidpunct}
{\mcitedefaultendpunct}{\mcitedefaultseppunct}\relax
\EndOfBibitem
\bibitem[Fromager(2025)]{Fromager2025indvElevel}
Fromager,~E. Ensemble Density Functional Theory of Ground and Excited Energy
  Levels. \emph{The Journal of Physical Chemistry A} \textbf{2025}, \emph{129},
  1143--1155, PMID: 39829255\relax
\mciteBstWouldAddEndPuncttrue
\mciteSetBstMidEndSepPunct{\mcitedefaultmidpunct}
{\mcitedefaultendpunct}{\mcitedefaultseppunct}\relax
\EndOfBibitem
\bibitem[Gould \latin{et~al.}(2025)Gould, Kronik, and
  Pittalis]{gould2025ensemblizationdensityfunctionaltheory}
Gould,~T.; Kronik,~L.; Pittalis,~S. "Ensemblization" of density functional
  theory. \emph{arXiv preprint physics.chem-ph} \textbf{2025},
  \url{https://arxiv.org/abs/2504.00547} (accessed 2025-08-18)\relax
\mciteBstWouldAddEndPuncttrue
\mciteSetBstMidEndSepPunct{\mcitedefaultmidpunct}
{\mcitedefaultendpunct}{\mcitedefaultseppunct}\relax
\EndOfBibitem
\bibitem[Deur and Fromager(2019)Deur, and Fromager]{deur2019ground}
Deur,~K.; Fromager,~E. Ground and excited energy levels can be extracted
  exactly from a single ensemble density-functional theory calculation.
  \emph{J. Chem. Phys.} \textbf{2019}, \emph{150}, 094106\relax
\mciteBstWouldAddEndPuncttrue
\mciteSetBstMidEndSepPunct{\mcitedefaultmidpunct}
{\mcitedefaultendpunct}{\mcitedefaultseppunct}\relax
\EndOfBibitem
\bibitem[Ullrich and Kohn(2001)Ullrich, and Kohn]{Ullrich2001EDFTdegen}
Ullrich,~C.~A.; Kohn,~W. Kohn-Sham Theory for Ground-State Ensembles.
  \emph{Phys. Rev. Lett.} \textbf{2001}, \emph{87}, 093001\relax
\mciteBstWouldAddEndPuncttrue
\mciteSetBstMidEndSepPunct{\mcitedefaultmidpunct}
{\mcitedefaultendpunct}{\mcitedefaultseppunct}\relax
\EndOfBibitem
\bibitem[Marut \latin{et~al.}(2020)Marut, Senjean, Fromager, and
  Loos]{marut2020weight}
Marut,~C.; Senjean,~B.; Fromager,~E.; Loos,~P.-F. Weight dependence of local
  exchange--correlation functionals in ensemble density-functional theory:
  double excitations in two-electron systems. \emph{Faraday Discuss.}
  \textbf{2020}, \emph{224}, 402--423\relax
\mciteBstWouldAddEndPuncttrue
\mciteSetBstMidEndSepPunct{\mcitedefaultmidpunct}
{\mcitedefaultendpunct}{\mcitedefaultseppunct}\relax
\EndOfBibitem
\bibitem[Sagredo and Burke(2018)Sagredo, and Burke]{sagredo2018can}
Sagredo,~F.; Burke,~K. Accurate double excitations from ensemble density
  functional calculations. \emph{J. Chem. Phys.} \textbf{2018}, \emph{149},
  134103\relax
\mciteBstWouldAddEndPuncttrue
\mciteSetBstMidEndSepPunct{\mcitedefaultmidpunct}
{\mcitedefaultendpunct}{\mcitedefaultseppunct}\relax
\EndOfBibitem
\bibitem[Amoyal \latin{et~al.}(2024)Amoyal, Kronik, and
  Gould]{Amoyal_2025_pEDFT}
Amoyal,~G.~S.; Kronik,~L.; Gould,~T. Perturbative ensemble density functional
  theory applied to charge transfer excitations. \emph{Journal of Physics:
  Condensed Matter} \textbf{2024}, \emph{37}, 095503\relax
\mciteBstWouldAddEndPuncttrue
\mciteSetBstMidEndSepPunct{\mcitedefaultmidpunct}
{\mcitedefaultendpunct}{\mcitedefaultseppunct}\relax
\EndOfBibitem
\bibitem[Senjean \latin{et~al.}(2015)Senjean, Knecht, Jensen, and
  Fromager]{senjean2015linear}
Senjean,~B.; Knecht,~S.; Jensen,~H. J.~{\Aa}.; Fromager,~E. Linear
  interpolation method in ensemble Kohn-Sham and range-separated
  density-functional approximations for excited states. \emph{Phys. Rev. A}
  \textbf{2015}, \emph{92}, 012518\relax
\mciteBstWouldAddEndPuncttrue
\mciteSetBstMidEndSepPunct{\mcitedefaultmidpunct}
{\mcitedefaultendpunct}{\mcitedefaultseppunct}\relax
\EndOfBibitem
\bibitem[Woods \latin{et~al.}(2019)Woods, Payne, and Hasnip]{Woods_2019scfdft}
Woods,~N.~D.; Payne,~M.~C.; Hasnip,~P.~J. Computing the self-consistent field
  in Kohn–Sham density functional theory. \emph{Journal of Physics: Condensed
  Matter} \textbf{2019}, \emph{31}, 453001\relax
\mciteBstWouldAddEndPuncttrue
\mciteSetBstMidEndSepPunct{\mcitedefaultmidpunct}
{\mcitedefaultendpunct}{\mcitedefaultseppunct}\relax
\EndOfBibitem
\bibitem[Levi \latin{et~al.}(2020)Levi, Ivanov, and Jónsson]{Levi_DOMOM_2020}
Levi,~G.; Ivanov,~A.~V.; Jónsson,~H. Variational Density Functional
  Calculations of Excited States via Direct Optimization. \emph{Journal of
  Chemical Theory and Computation} \textbf{2020}, \emph{16}, 6968--6982, PMID:
  33064484\relax
\mciteBstWouldAddEndPuncttrue
\mciteSetBstMidEndSepPunct{\mcitedefaultmidpunct}
{\mcitedefaultendpunct}{\mcitedefaultseppunct}\relax
\EndOfBibitem
\bibitem[Schmerwitz \latin{et~al.}(2023)Schmerwitz, Levi, and
  Jónsson]{OODFT_GM_2023}
Schmerwitz,~Y. L.~A.; Levi,~G.; Jónsson,~H. Calculations of Excited Electronic
  States by Converging on Saddle Points Using Generalized Mode Following.
  \emph{Journal of Chemical Theory and Computation} \textbf{2023}, \emph{19},
  3634--3651, PMID: 37283439\relax
\mciteBstWouldAddEndPuncttrue
\mciteSetBstMidEndSepPunct{\mcitedefaultmidpunct}
{\mcitedefaultendpunct}{\mcitedefaultseppunct}\relax
\EndOfBibitem
\bibitem[Schmerwitz \latin{et~al.}(2025)Schmerwitz, Selenius, and
  Levi]{schmerwitz2025freezeandreleasedirectoptimizationmethod}
Schmerwitz,~Y. L.~A.; Selenius,~E.; Levi,~G. Freeze-and-release direct
  optimization method for variational calculations of excited electronic
  states. \emph{arXiv preprint physics.chem-ph} \textbf{2025},
  \url{https://arxiv.org/abs/2501.18568} (accessed 2025-08-18)\relax
\mciteBstWouldAddEndPuncttrue
\mciteSetBstMidEndSepPunct{\mcitedefaultmidpunct}
{\mcitedefaultendpunct}{\mcitedefaultseppunct}\relax
\EndOfBibitem
\bibitem[Olsen and Jo/rgensen(1985)Olsen, and Jo/rgensen]{LRMCSCF1985}
Olsen,~J.; Jo/rgensen,~P. Linear and nonlinear response functions for an exact
  state and for an MCSCF state. \emph{The Journal of Chemical Physics}
  \textbf{1985}, \emph{82}, 3235--3264\relax
\mciteBstWouldAddEndPuncttrue
\mciteSetBstMidEndSepPunct{\mcitedefaultmidpunct}
{\mcitedefaultendpunct}{\mcitedefaultseppunct}\relax
\EndOfBibitem
\bibitem[Heid \latin{et~al.}(2018)Heid, Hunt, and Schröder]{C7CP08549D}
Heid,~E.; Hunt,~P.~A.; Schröder,~C. Evaluating excited state atomic
  polarizabilities of chromophores. \emph{Phys. Chem. Chem. Phys.}
  \textbf{2018}, \emph{20}, 8554--8563\relax
\mciteBstWouldAddEndPuncttrue
\mciteSetBstMidEndSepPunct{\mcitedefaultmidpunct}
{\mcitedefaultendpunct}{\mcitedefaultseppunct}\relax
\EndOfBibitem
\bibitem[Mosquera \latin{et~al.}(2016)Mosquera, Chen, Ratner, and
  Schatz]{SLRTDDFT2016}
Mosquera,~M.~A.; Chen,~L.~X.; Ratner,~M.~A.; Schatz,~G.~C. Sequential double
  excitations from linear-response time-dependent density functional theory.
  \emph{The Journal of Chemical Physics} \textbf{2016}, \emph{144},
  204105\relax
\mciteBstWouldAddEndPuncttrue
\mciteSetBstMidEndSepPunct{\mcitedefaultmidpunct}
{\mcitedefaultendpunct}{\mcitedefaultseppunct}\relax
\EndOfBibitem
\bibitem[Mosquera \latin{et~al.}(2021)Mosquera, Jones, Kang, Ratner, and
  Schatz]{SLRTDDFT2021}
Mosquera,~M.~A.; Jones,~L.~O.; Kang,~G.; Ratner,~M.~A.; Schatz,~G.~C. Second
  Linear Response Theory and the Analytic Calculation of Excited-State
  Properties. \emph{The Journal of Physical Chemistry A} \textbf{2021},
  \emph{125}, 1093--1102, PMID: 33497573\relax
\mciteBstWouldAddEndPuncttrue
\mciteSetBstMidEndSepPunct{\mcitedefaultmidpunct}
{\mcitedefaultendpunct}{\mcitedefaultseppunct}\relax
\EndOfBibitem
\bibitem[Senjean and Fromager(2018)Senjean, and Fromager]{senjean2018unified}
Senjean,~B.; Fromager,~E. Unified formulation of fundamental and optical gap
  problems in density-functional theory for ensembles. \emph{Phys. Rev. A}
  \textbf{2018}, \emph{98}, 022513\relax
\mciteBstWouldAddEndPuncttrue
\mciteSetBstMidEndSepPunct{\mcitedefaultmidpunct}
{\mcitedefaultendpunct}{\mcitedefaultseppunct}\relax
\EndOfBibitem
\bibitem[Perdew and Levy(1983)Perdew, and Levy]{perdew1983physical}
Perdew,~J.~P.; Levy,~M. Physical content of the exact Kohn-Sham orbital
  energies: band gaps and derivative discontinuities. \emph{Phys. Rev. Lett.}
  \textbf{1983}, \emph{51}, 1884\relax
\mciteBstWouldAddEndPuncttrue
\mciteSetBstMidEndSepPunct{\mcitedefaultmidpunct}
{\mcitedefaultendpunct}{\mcitedefaultseppunct}\relax
\EndOfBibitem
\bibitem[Hodgson \latin{et~al.}(2021)Hodgson, Wetherell, and
  Fromager]{PRA21_Hodgson_exact_Nc-eDFT_1D}
Hodgson,~M. J.~P.; Wetherell,~J.; Fromager,~E. Exact exchange-correlation
  potentials for calculating the fundamental gap with a fixed number of
  electrons. \emph{Phys. Rev. A} \textbf{2021}, \emph{103}, 012806\relax
\mciteBstWouldAddEndPuncttrue
\mciteSetBstMidEndSepPunct{\mcitedefaultmidpunct}
{\mcitedefaultendpunct}{\mcitedefaultseppunct}\relax
\EndOfBibitem
\bibitem[Hellgren and Gross(2012)Hellgren, and Gross]{Hellgren12_Effect}
Hellgren,~M.; Gross,~E. K.~U. {Effect of discontinuities in Kohn-Sham-based
  chemical reactivity theory}. \emph{The Journal of Chemical Physics}
  \textbf{2012}, \emph{136}, 114102\relax
\mciteBstWouldAddEndPuncttrue
\mciteSetBstMidEndSepPunct{\mcitedefaultmidpunct}
{\mcitedefaultendpunct}{\mcitedefaultseppunct}\relax
\EndOfBibitem
\bibitem[Kraisler and Kronik(2013)Kraisler, and Kronik]{kraisler2013piecewise}
Kraisler,~E.; Kronik,~L. Piecewise linearity of approximate density functionals
  revisited: implications for frontier orbital energies. \emph{Phys. Rev.
  Lett.} \textbf{2013}, \emph{110}, 126403\relax
\mciteBstWouldAddEndPuncttrue
\mciteSetBstMidEndSepPunct{\mcitedefaultmidpunct}
{\mcitedefaultendpunct}{\mcitedefaultseppunct}\relax
\EndOfBibitem
\bibitem[Nagy(1995)]{Nagy_ensAC}
Nagy,~A. Coordinate scaling and adiabatic connection formula for ensembles of
  fractionally occupied excited states. \emph{Int. J. Quantum Chem.}
  \textbf{1995}, \emph{56}, 225--228\relax
\mciteBstWouldAddEndPuncttrue
\mciteSetBstMidEndSepPunct{\mcitedefaultmidpunct}
{\mcitedefaultendpunct}{\mcitedefaultseppunct}\relax
\EndOfBibitem
\bibitem[Daas \latin{et~al.}(2025)Daas, Crisostomo, and
  Burke]{daas2025ensembletimedependentdensityfunctional}
Daas,~K.~J.; Crisostomo,~S.; Burke,~K. Ensemble Time-Dependent Density
  Functional Theory. \emph{arXiv preprint physics.chem-ph} \textbf{2025},
  \url{https://arxiv.org/abs/2507.19464} (accessed 2025-08-18)\relax
\mciteBstWouldAddEndPuncttrue
\mciteSetBstMidEndSepPunct{\mcitedefaultmidpunct}
{\mcitedefaultendpunct}{\mcitedefaultseppunct}\relax
\EndOfBibitem
\end{mcitethebibliography}

\end{document}